\RequirePackage{fix-cm}
\documentclass[twocolumn,epjc3]{svjour3}
\smartqed  
\RequirePackage{graphicx}
\RequirePackage{cite} \RequirePackage{amsmath,amsfonts,amssymb}
\RequirePackage{url} \RequirePackage[british]{babel}
\RequirePackage{hyphenat}

\journalname{Eur. Phys. J. C}

\begin{document}

\title{Assessment of backgrounds of the ANAIS experiment for dark matter direct
detection}


\author{J.~Amar\'{e}\thanksref{addr1,addr2}
        \and S.~Cebri\'{a}n\thanksref{e1,addr1,addr2}
        \and C.~Cuesta\thanksref{addr1,addr2,addr3}
        \and E.~Garc\'{i}a\thanksref{addr1,addr2}
        \and M.~Mart\'{i}nez\thanksref{addr1,addr2,addr4}
        \and M.A.~Oliv\'{a}n\thanksref{addr1,addr2}
        \and Y.~Ortigoza\thanksref{addr1,addr2}
        \and A.~Ortiz de Sol\'{o}rzano\thanksref{addr1,addr2}
        \and J.~Puimed\'{o}n\thanksref{addr1,addr2}
        \and M.L.~Sarsa\thanksref{addr1,addr2}
        \and J.A.~Villar\thanksref{addr1,addr2}
        \and P.~Villar\thanksref{addr1,addr2}
        }

\thankstext{e1}{e-mail: scebrian@unizar.es}


\institute{Laboratorio de F\'{i}sica Nuclear y Astropart\'{i}culas,
Universidad de Zaragoza, Calle Pedro Cerbuna 12, 50009 Zaragoza,
Spain \label{addr1}
           \and
           Laboratorio Subterr\'{a}neo de Canfranc, Paseo de los Ayerbe s/n,
22880 Canfranc Estaci\'{o}n, Huesca, Spain\label{addr2}
           \and
           \emph{Present Address:} Department of Physics, Center for Experimental Nuclear Physics and Astrophysics, University of Washington, Seattle, WA, USA\label{addr3}
           \and
           \emph{Present Address:} Universit\`{a} di Roma La Sapienza, Piazzale Aldo Moro 5, 00185 Roma, Italy\label{addr4}
}

\date{Received: date / Accepted: date}

\maketitle

\begin{abstract}
A large effort has been carried out to characterize the background
of sodium iodide crystals within the ANAIS (Annual modulation with
NaI Scintillators) project. In this paper, the background models
developed for three 12.5-kg NaI(Tl) scintillators produced by Alpha
Spectra Inc. and operated at the Canfranc Underground Laboratory are
presented together with an evaluation of the background prospects
for the full experiment. Measured spectra from threshold to high
energy in different conditions are well described by the models
based on quantified activities. At the region of interest, crystal
bulk contamination is the dominant background source. Contributions
from $^{210}$Pb, $^{40}$K, $^{22}$Na and $^{3}$H are the most
relevant. Those from $^{40}$K and $^{22}$Na could be efficiently
suppressed thanks to anticoincidence operation in a crystals matrix
or inside a Liquid Scintillator Veto (LSV), while that from
$^{210}$Pb has been reduced by improving crystal production methods
and $^{3}$H production could be reduced by shielding against cosmic
rays during production. Assuming the activities of the last
characterized detector, for nine crystals with a total mass of
112.5~kg the expected background rate is 2.5~counts/(keV~kg~d) in
the region from 1 to 4~keV, which could be reduced at
1.4~counts/(keV~kg~d) by using a LSV.
\end{abstract}

\section{Introduction}

Sodium iodide crystals doped with Tl have been widely used as
radiation detectors and, in particular, they have been applied in
the direct search of galactic dark matter particles for a long time
\cite{fushimi99,sarsa97,bernabei99,gerbier99,naiad}. Among the
several experimental approaches using NaI(Tl) detectors, DAMA/ LIBRA
is the most relevant, having reported the observation of a
modulation compatible with that expected for galactic halo WIMPs
with a large statistical significance \cite{bernabei13}. Results
obtained with other target materials and detection techniques (like
those from CDMS~\cite{cdmsresults}, CRESST~\cite{cresstresults},
EDELWEISS~\cite{edelweissresults}, KIMS~\cite{kimsresults},
LUX~\cite{luxresults}, PICO~\cite{picoresults} or
XENON~\cite{xenonresults} collaborations) have been ruling out for
years the most plausible compatibility scenarios. The ANAIS (Annual
modulation with NaI Scintillators) project \cite{anais15} is
intended to search for dark matter annual modulation with ultrapure
NaI(Tl) scintillators at the Canfranc Underground Laboratory (LSC)
in Spain. The aim is to provide a model-independent confirmation of
the annual modulation positive signal reported by DAMA/LIBRA using
the same target and technique, but different experimental conditions
affecting systematics. Projects like DM-Ice \cite{dmice}, KIMS
\cite{kims} and SABRE \cite{sabre} also envisage the use of large
masses of NaI(Tl) for dark matter searches.

ANAIS aims at the study of the annual modulation signal using a
NaI(Tl) mass of 112.5~kg at the LSC. To confirm the DAMA/LIBRA
results, ANAIS detectors should be comparable to those of DAMA/LIBRA
in terms of energy threshold and radioactive background: energy
threshold lower than 2~keVee\footnote{Electron equivalent energy.}
and background at 1-2~counts/(keV~kg~d) in the region of interest
(RoI) below 6~keVee. Several prototypes have been developed and
operated at LSC using BICRON and Saint-Gobain crystals; all of them
were disregarded due to an unacceptable K content in the crystal at
the level of hundreds of ppb. Among them, the so-called ANAIS--0
detector \cite{anaisap,anaisijmpa,anaisom,anaisepjc}, a 9.6~kg
Saint-Gobain crystal similar to those of DAMA experiment, has to be
highlighted because its successful background model \cite{anaisap}
has been the starting point for the present work. Some other
interesting results, as very slow scintillation in NaI(Tl)
\cite{anaisom} or an anomalous fast event population attributable to
quartz light emission \cite{anaisom2} were also obtained from first
prototypes. The main challenge for ANAIS has been the achievement of
the required low background level, being contaminations in the bulk
of the crystal the dominant contribution in the RoI.

The new prototypes built by Alpha Spectra Inc. consist of 12.5~kg
NaI(Tl) crystals, housed in OFE (Oxygen Free Electronic) copper and
coupled through quartz windows to two Hamamatsu photomultipliers
(PMTs) at the LSC clean room in a second step; they have been fully
tested and characterized at the LSC since the end of 2012, obtaining
very promising results.

In the full ANAIS experiment, the total NaI(Tl) active mass will be
divided into a number of such modules: nine of them, accounting for
112.5~kg, will be set-up at LSC along 2016. The shielding for the
experiment will consist of 10~cm of archaeological lead, 20~cm of
low activity lead, 40~cm of neutron moderator, an anti-radon box (to
be continuously flushed with radon-free air) and an active muon veto
system made up of plastic scintillators designed to cover top and
sides of the whole ANAIS set-up. The hut that will house the
experiment at the hall B of LSC (under 2450~m.w.e.) is already
operative, shielding materials and electronic chain components are
prepared for mounting \cite{anaisricap}. Different PMT models were
tested in order to choose the best option in terms of light
collection and background \cite{Clarathesis}. The Hamamatsu
R12669SEL2 was selected and all the required units are available at
the LSC.

The construction of reliable background models is essential for
experiments demanding ultra low background conditions since they
provide guidance and constraints for design and for analyzing any
possible systematics and allow robust estimations of the experiment
sensitivity (see some recent examples at
\cite{xenon100,zeplin,edelweiss,gerda,lux,exo}). It is worth noting
that the reliability of this kind of studies depends on three
important aspects: an accurate assay of background sources, a
careful computation of their contribution to the experiment
(typically made by Monte Carlo simulation) and continuous validation
of the obtained results against experimental data, which will be
stronger if data in different experimental conditions and energy
ranges are considered. On the other hand, a complete understanding
of the DAMA/LIBRA background at low energy has not yet been achieved
and some open questions remain \cite{kudry10,bernabeianswer}.
Therefore, a careful analysis and quantification of the different
background components in the ANAIS prototypes produced by Alpha
Spectra was undertaken and is presented here. The quantification of
cosmogenic radionuclide production and its effects in NaI(Tl)
crystals using data from the two first prototypes (D0 and D1, see
section~\ref{secanais25}) have been specifically studied at
\cite{anaisjcap}.

The structure of the article is the following. The experimental
set-ups of the detectors and the measurements taken are described in
section~\ref{setups}. Sections~\ref{sources} and \ref{modeling}
present the background sources considered and the details of their
simulation. The quantified background contributions and the
comparison with data at different energy ranges and experimental
conditions are discussed in section~\ref{comparison}. Finally, the
background projections for the full ANAIS experiment based on the
obtained results with the first modules are shown in
section~\ref{projections}, while conclusions are summarized in
section~\ref{conclusions}.

\section{Experimental set-ups} \label{setups}

Two prototypes of 12.5~kg mass (named D0 and D1), made by Alpha
Spectra (AS) Inc., CO (US), with ultrapure NaI powder, took data at
the LSC from December 2012 to March 2015; we will refer in the
following to this set-up as ANAIS--25 \cite{anaisnima}.
Its main goals were to measure the crystal internal contamination,
determine light collection, fine tune the data acquisition and test
the filtering and analysis protocols.

\begin{figure*}
\centering
 \includegraphics[width=0.45\textwidth]{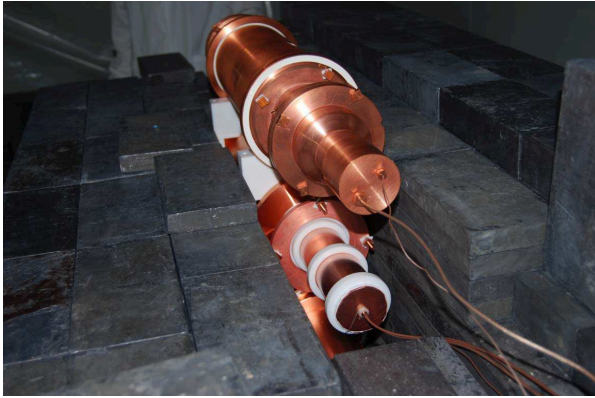}
 \includegraphics[width=0.45\textwidth]{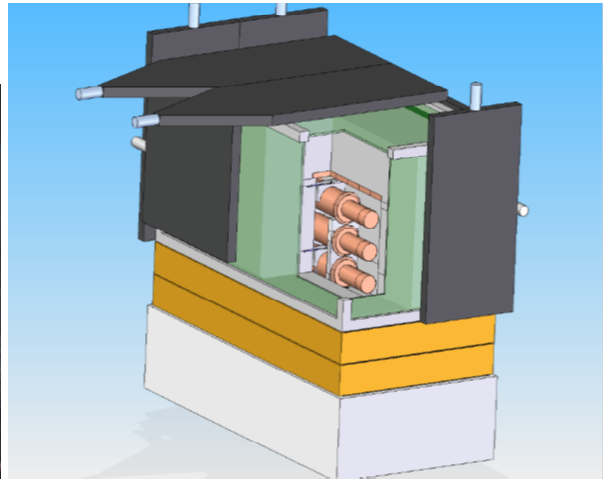}
 \caption{Picture (left) and design (right) of the ANAIS--37 set-up at LSC.}
  \label{anais37setup}
\end{figure*}

The ANAIS--37 set-up \cite{anaislrt,anaispatras} combines the
ANAIS--25 modules with a new one (named D2) also built by AS, using
improved protocols in order to prevent radon contamination and
WIMPScint-II grade powder. The crystal was received on the 6$^{th}$
of March, 2015 and
data taking started five days later. Data considered in this work
were taken from March to September 2015. The new module (D2) was
placed in between the two ANAIS--25 modules (D0 and D1) to maximize
the coincidence efficiency for the potassium determination (see
figure~\ref{anais37setup}). The main goal of ANAIS--37 set-up was to
characterize the new D2 module, in particular, to evaluate the
reduction of $^{210}$Pb contamination, to check the content of
$^{40}$K and $^{238}$U and $^{232}$Th chains and to assess also its
general performance.

The three AS modules are cylindrical, 4.75'' diameter and 11.75''
length, with synthetic quartz windows for PMTs coupling; they were
encapsulated following similar protocols and using the same
materials. Hamamatsu R12669SEL2 PMTs were coupled at LSC clean room
for the three detectors. A Mylar window in the lateral face allows
for low energy calibration. The shielding in both set-ups consisted
of 10~cm of archaeological lead plus 20~cm of low activity lead
shielding inside a radon exclusion box at LSC. An active muon veto
system made up of plastic scintillators covered the top and sides of
the shielding, first the coverage was only partial, but at the end
of the ANAIS--37 data taking period the veto was fully operative.

Concerning the data acquisition system of these set-ups, each PMT
charge output signal is separately processed for obtaining trigger,
pulse shape digitization and energy at different ranges. Triggering
is done by the coincidence (logical AND) of the two PMT signals of
any detector at photoelectron level in a 200~ns window, enabling
digitization and conversion of the two signals. The building of the
spectra is done by software (off-line) by adding the signals from
both PMTs, and Pulse Shape Analysis is applied in order to select
bulk scintillation events in the NaI crystals and to distinguish
alpha interactions from beta/gamma ones. Data considered here
correspond to the low and high energy regions, below 200~keV and up
to 3~MeV, respectively. Filtering protocols for PMT noise similar to
those described at \cite{anaisepjc} for ANAIS--0 prototype but
optimized for these new detectors have been applied, and a threshold
at 1~keVee has been considered in the following, although work is
still ongoing in order to improve the filtering and the acceptance
efficiency estimate, probably underestimated at present
\cite{anais15}.

Regarding the response of the detectors, it must be remarked the
outstanding light collection measured for the three AS modules, at
the level of $\sim$15~phe/keV \cite{MAthesis}, which is a factor of
2 larger than that determined for the best DAMA/LIBRA detectors
\cite{bernabei2008}. This much higher light output has a direct
impact in both resolution and energy threshold, but it also allows
to improve strongly the signal vs noise filtering down to the
threshold and hence, reduce analysis uncertainties.

\section{Background contributions} \label{sources}

The background sources considered for the ANAIS prototypes include
activities from crystals as well as from external components. The
latter have been mainly directly assayed by HPGe spectrometry at
LSC. Table~\ref{extcon} summarizes the measured activities (or
derived upper limits) of the components used in the set-up and taken
into consideration. Every PMT unit to be used in the full experiment
has been already screened, finding compatible levels of activity
among them; values quoted in table~\ref{extcon} correspond to the
six units used in the ANAIS--25 and ANAIS--37 set-ups and to the
mean from all screened units. For copper and quartz windows values
are as for ANAIS--0 prototype \cite{anaisap}. For radon content in
the air filling the inner volume of the shielding, there is no
direct measurement. Radon in the laboratory air is being
continuously monitored, and the inner volume of the shielding is
flushed with boil-off nitrogen, to guarantee its radon-free quality.
A value for the radon content in the inner volume air of about one
hundredth of the external air radon content has been assumed in our
background model (0.6~Bq/m$^{3}$), compatible with the absence of
lines coming from radon daughter isotopes in the measured
background. Contribution from fast neutrons and environmental gamma
background has been also estimated, being negligible at the present
level of sensitivity. Contribution from muons interacting in the
crystal (and other muon related events) can be vetoed by the
coincidence with a signal in the plastic scintillators covering the
shielding and then, it has not been considered in our background
model. Although the active veto was not in operation during the data
taking this work refers to, the muon induced background is
negligible \cite{anaisepjc}.

\begin{table*}
\caption{Activity of the external components (outside crystal) of
the ANAIS prototypes considered as background sources. Except for
the inner volume air, the values have been measured by HPGe
spectrometry performed at the LSC. Upper limits are given at 95\%
C.L.}
\label{extcon}
\begin{center}
\begin{tabular}{lcccccc}
\hline\noalign{\smallskip} Component &  Unit  &  $^{40}$K &
$^{232}$Th &  $^{238}$U  & $^{226}$Ra & Others
 \\ \noalign{\smallskip}\hline\noalign{\smallskip}

PMTs (R12669SEL2) & mBq/PMT & 97$\pm$19 & 20$\pm$2 & 128$\pm$38 &
84$\pm$3 & \\
& & 133$\pm$13 &  20$\pm$2 &  150$\pm$34  & 88$\pm$3 & \\
& & 108$\pm$29 & 21$\pm$3 & 161$\pm$58 & 79$\pm$56 & \\
& & 95$\pm$24 & 22$\pm$2 & 145$\pm$29 & 88$\pm$4 & \\
& & 136$\pm$26 & 18$\pm$2 & 187$\pm$58 & 59$\pm$3 \\
& & 155$\pm$36 & 20$\pm$3 & 144$\pm$33 & 89$\pm$5 \\
 mean activity all units & mBq/PMT & 111$\pm$5 &  20.7$\pm$0.5  &
157$\pm$8 & 82.5$\pm$0.8 \\ \hline

Copper encapsulation & mBq/kg & $<$4.9 &  $<$1.8 &  $<$62 & $<$0.9
&$^{60}$Co: $<$0.4 \\ \hline

Quartz windows &  mBq/kg & $<$12 & $<$2.2 & $<$100 & $<$1.9 & \\
\hline

Silicone pads & mBq/kg & $<$181 &  $<$34 & &  51$\pm$7 & \\ \hline

Archaelogical lead & mBq/kg & &  $<$0.3 & $<$0.2 & & $^{210}$Pb:
$<$20 \\ \hline

Inner volume air & Bq/m$^{3}$ &   & & & &  $^{222}$Rn: 0.6
\\ \noalign{\smallskip}\hline
\end{tabular}
\end{center}
\end{table*}

Concerning the background from the NaI(Tl) crystals, the activity of
the most relevant radionuclides has been directly measured for our
crystals applying different techniques \cite{anaisnima}. Bulk
$^{40}$K content is estimated by searching for the coincidences
between 3.2~keV energy deposition in one detector (following EC) and
the 1461~keV gamma line escaping from it and being fully absorbed in
other detector \cite{anaisijmpa}; efficiency of the coincidence is
estimated by Monte Carlo simulation. The activities from $^{210}$Pb
and $^{232}$Th and $^{238}$U chains have been determined on the one
hand, by quantifying Bi/Po and alpha-alpha sequences, and on the
other, by comparing the total alpha rate determined through pulse
shape analysis with the low energy depositions attributable to
$^{210}$Pb, which are fully compatible. The results obtained for D0
and D1 detectors using ANAIS--25 data (see table~\ref{intcon}) gave
a moderate contamination of $^{40}$K, above the initial goal of
ANAIS (20~ppb of K) but acceptable, a low content of $^{232}$Th and
$^{238}$U chains but a high activity of $^{210}$Pb at the mBq/kg
level. The origin of such contamination was identified and addressed
by Alpha Spectra. According to preliminary results from ANAIS--37
corresponding to 168.1~days of live-time and shown in table
\ref{intcon}, an average total alpha activity of
0.70$\pm$0.10~mBq/kg has been observed for the new module D2, which
is a factor 5 lower than the alpha activity in ANAIS--25 modules;
therefore, it can be concluded that effective reduction of Rn
entrance in the growing and/or purification at Alpha Spectra has
been achieved and is expected to improve for next prototypes. The
potassium content of D2 was analyzed using the same technique
applied to previous prototypes, obtaining a value compatible with
those obtained for D0 and D1. The potassium concentration and alpha
activity reported by the KIMS collaboration for a similar crystal
produced also by Alpha Spectra from WIMPScint-II material
\cite{kims2} are of the same order than those measured for D2.

\begin{table*}
\caption{Measured activity in NaI(Tl) crystals for D0, D1 and D2
detectors using ANAIS--25 and ANAIS--37 data and combining different
analysis techniques.} \label{intcon}
\begin{center}
\begin{tabular}{lccccc}
\hline\noalign{\smallskip} Detector &  Unit  &  $^{40}$K &
$^{232}$Th &  $^{238}$U  & $^{210}$Pb
 \\ \noalign{\smallskip}\hline\noalign{\smallskip}
D0, D1 & mBq/kg & 1.4$\pm$0.2 (D0)  & (4$\pm$1) 10$^{-3}$
&(10$\pm$2) 10$^{-3}$ & 3.15$\pm$0.10
\\ 
& & 1.1$\pm$0.2 (D1) & & & \\
D2 & mBq/kg & 1.1$\pm$0.2 & (0.7$\pm$0.1) 10$^{-3}$ & (2.7$\pm$0.2)
10$^{-3}$& 0.70$\pm$0.10
\\ \noalign{\smallskip}\hline
\end{tabular}
\end{center}
\end{table*}

Thanks to the very good detector response and the prompt data taking
starting after storing the detectors underground, a detailed study
of cosmogenic radionuclide production in NaI(Tl) has been performed
from ANAIS--25 data \cite{anaisjcap}. The initial activity, $A_{0}$,
corresponding to the moment of storing crystals deep underground at
LSC was deduced, studying the exponential decay of the identifying
signature produced by each isotope. The crystal growing and detector
manufacture took place at Alpha Spectra facilities in Grand
Junction, CO (US) and detectors were taken from US to Spain by boat.
The production of some induced I, Te and Na isotopes was well
characterized and it was considered as a background source of the
detectors too. Table~\ref{cosact} shows the list of all identified
products and their half-lives, together with the measured initial
activities. $^{22}$Na could be specially worrisome for dark matter
searches because the binding energy of the K-shell of its daughter
Ne is 0.87~keV, falling the corresponding energy deposition in the
RoI, and having a long enough half-life to compromise the first
years of data taking. A direct estimate of $^{22}$Na activity in D2
crystal was carried out by analyzing coincidences\footnote{Data
corresponding to 111.4~days from a special set-up from October 2015
to February 2016 with only D0 and D2 detectors were used. In
particular, profiting from the reduced cosmogenics in this period,
D2 spectrum in coincidence with 1274.5~keV depositions in D0 was
analyzed. It is worth noting that the $^{22}$Na initial activity in
D0 deduced from the analogue analysis is in perfect agreement with
the first estimate in \cite{anaisjcap}.};
the obtained value for the initial activity,
$A_{0}$=(70.2$\pm$3.9)~kg$^{-1}$d$^{-1}$, is more than a factor of
two lower than the one deduced for D0 and D1 detectors. This result
is compatible with a lower time of exposure to cosmic rays, taking
into account the $^{22}$Na half-life, longer than that corresponding
to I and Te products. As it will be discussed later, there are hints
of the production of other isotopes like $^{3}$H, $^{109}$Cd and
$^{113}$Sn in the NaI(Tl) crystals, even if they could not be
directly identified in the first analysis of cosmogenic activation
presented in \cite{anaisjcap}. Apart from this, $^{129}$I can be
present in the NaI crystals; it can be produced either as residual
product of uranium spontaneous fission, or by cosmic rays reactions,
having a broad range of activity values in iodine compounds
depending on the ore origin. Due to its long lifetime and the
difficulty to disentangle its signal from other emissions, the
amount of $^{129}$I in ANAIS--25 crystals could not be quantified.
To take it into account in the background model its concentration
was assumed to be the same as estimated by DAMA/LIBRA
($^{129}$I/$^{nat}$I = (1.7$\pm$0.1) 10$^{-13}$)
\cite{bernabei2008}, corresponding to an activity of 0.94~mBq/kg.

\begin{table}
\caption{Measured initial activities underground ($A_{0}$) for the
identified cosmogenic isotopes in NaI(Tl) crystals using ANAIS--25
data \cite{anaisjcap}. Half-lives of the products are also indicated
\cite{ddep}.} \label{cosact}
\begin{center}
\begin{tabular}{lcc}
\hline\noalign{\smallskip} Isotope & $T_{1/2}$ & $A_{0}$ \\
& (days)&  (kg$^{-1}$d$^{-1}$) \\
\noalign{\smallskip}\hline\noalign{\smallskip}
$^{126}$I & 12.93$\pm$0.05  &  430$\pm$37  \\
$^{125}$I  & 59.407$\pm$0.009 &  621.8$\pm$1.6  \\
$^{127m}$Te  & 107$\pm$4 & 32.1$\pm$0.8 \\
$^{125m}$Te  & 57.40$\pm$0.15 &  79.1$\pm$0.8 \\
$^{123m}$Te   & 119.3$\pm$0.1 & 100.8$\pm$0.8   \\
$^{121m}$Te  & 154$\pm$7 & 76.9$\pm$0.8 \\
$^{121}$Te & 19.16$\pm$0.05 & 110$\pm$12  \\
$^{22}$Na & (2.6029$\pm$0.0008) y   & 159.7$\pm$4.9 \\
\noalign{\smallskip}\hline
\end{tabular}
\end{center}
\end{table}

\section{Background modeling} \label{modeling}

\begin{figure*}
\centering
 \includegraphics[height=0.27\textheight]{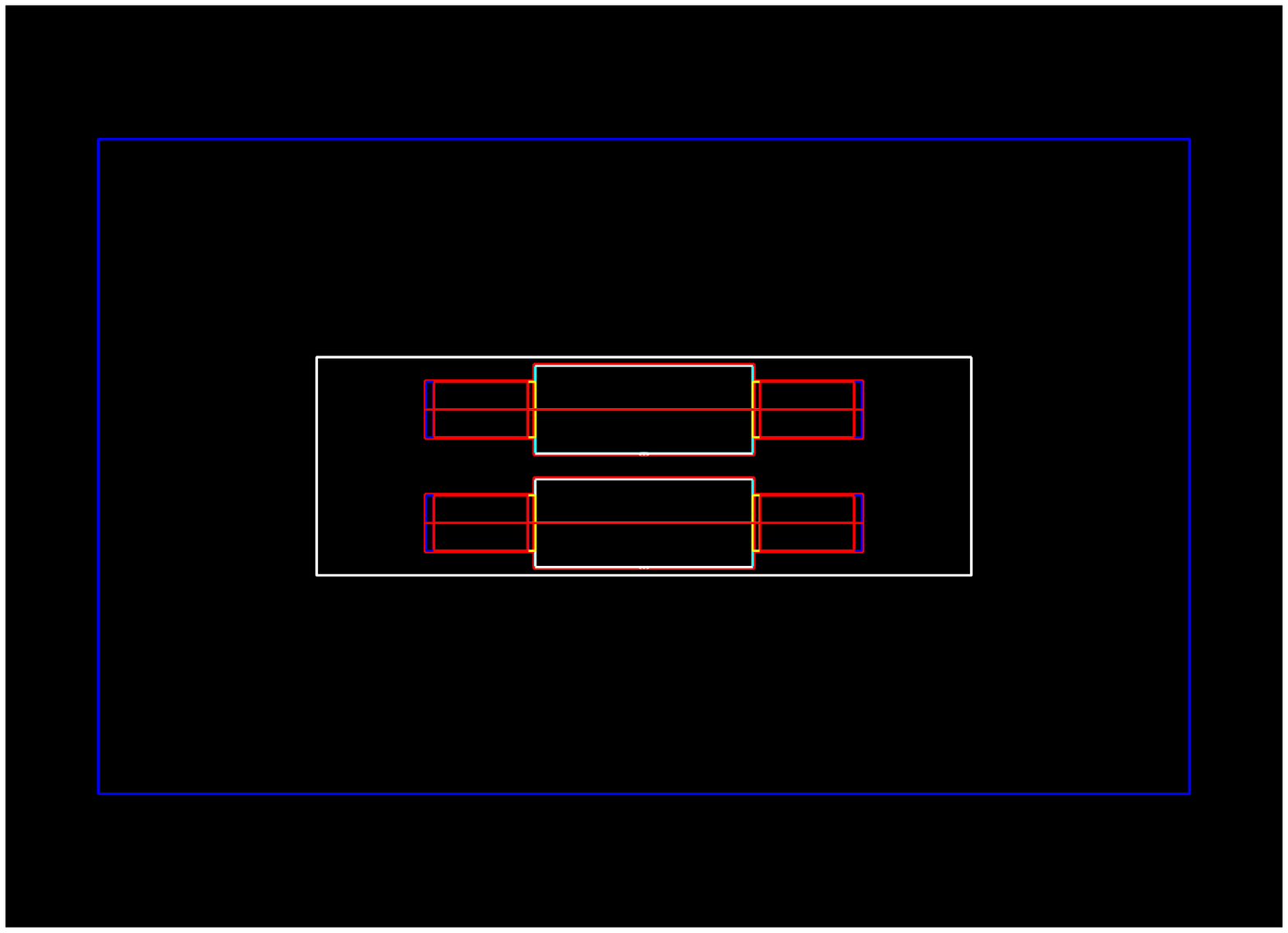} 
 \includegraphics[height=0.27\textheight]{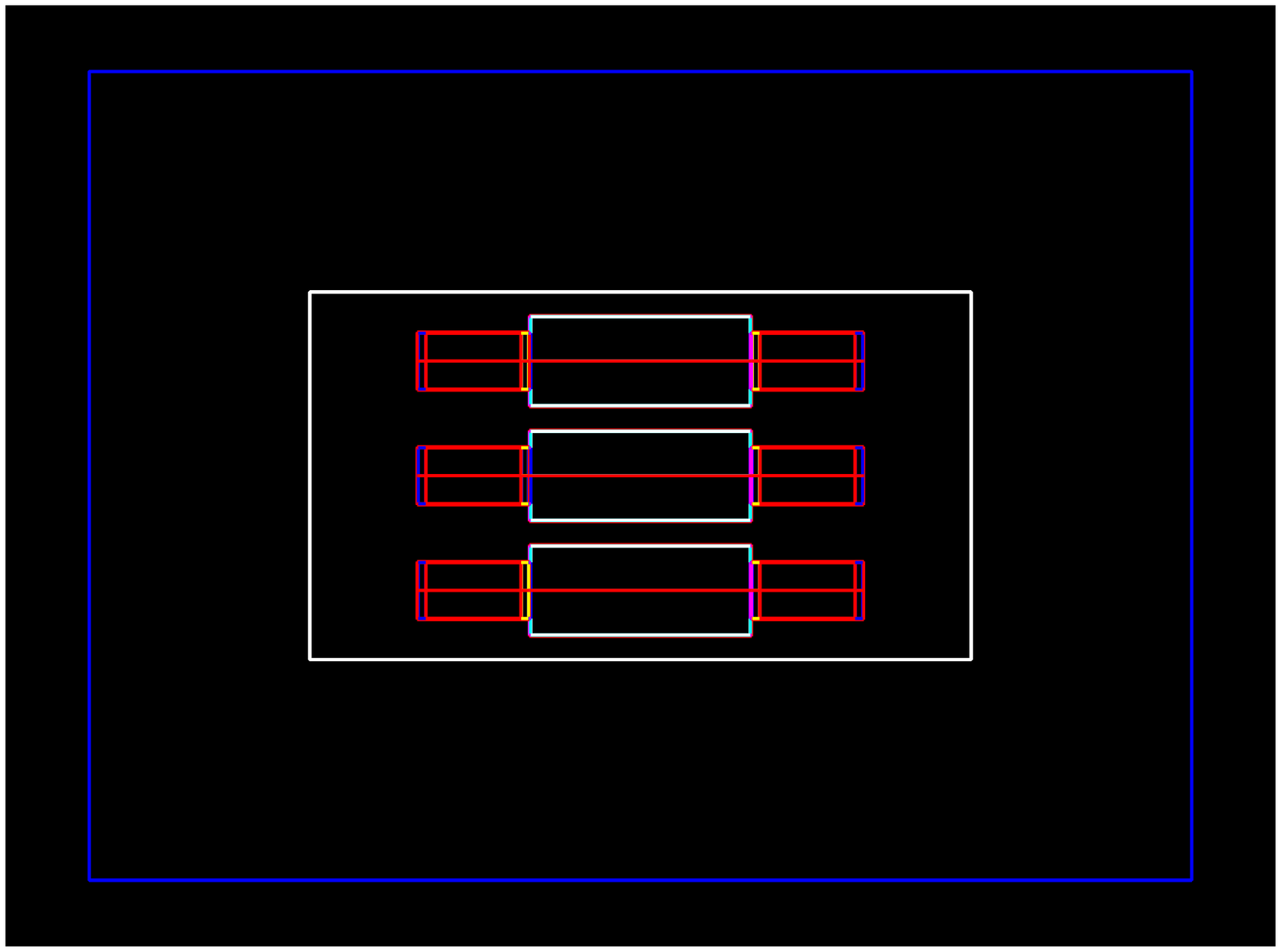} 
 \caption{Geometry of the ANAIS--25 (left) and ANAIS--37 (right) set-ups implemented in the Geant 4 simulations.}
  \label{geometrys}
\end{figure*}

The contribution of all the background sources described in section
\ref{sources} to the background levels of the ANAIS prototypes has
been simulated by Monte Carlo using the Geant4 package
\cite{geant4}, as done in \cite{anaisap}. A detailed description of
the set-ups was implemented including the lead shielding and
detectors, considering NaI crystal, teflon wrapping, copper
encapsulation with the Mylar window, silicone pads, quartz windows,
PMTs, bases and copper enclosure; figure~\ref{geometrys} shows the
views of the Geant4 geometry for ANAIS--25 and ANAIS--37 set-ups.
The Geant4 Radioactive Decay Module was used for simulating decays,
after checking carefully the energy conservation in the decay of all
the considered isotopes. The low energy models based on Livermore
data libraries were considered for the physical processes of
$\alpha$, $\beta$ and $\gamma$ emissions. Uniformly distributed bulk
contamination in the components was assumed and activities (or
derived upper limits) given in tables \ref{extcon}, \ref{intcon} and
\ref{cosact} considered. For each simulated event, defined
considering an energy integration time of 1 $\mu$s, the energy
deposited at each detector by different types of particles has been
recorded separately in order to build afterwards the energy
spectrum, filtering alpha deposits above 2.5~MeV (as it can be made
in real data by pulse shape analysis) and correcting each component
with the corresponding Relative Scintillation Efficiency
Factor\footnote{A constant value of 0.6 has been taken as relative
efficiency factor for alpha particles in the building of the
electron equivalent energy spectra. Energy from nuclear recoils is
neglected.}. Production of scintillation at the NaI(Tl) crystals and
the subsequent light collection have not been simulated here. Energy
spectra at different conditions have been constructed to allow
direct comparison to data obtained from detectors.

\section{Background contributions and comparison with data}\label{comparison}

\subsection{ANAIS--25 detectors}
\label{secanais25}

\begin{figure*}
\centering
 \includegraphics[width=0.45\textwidth]{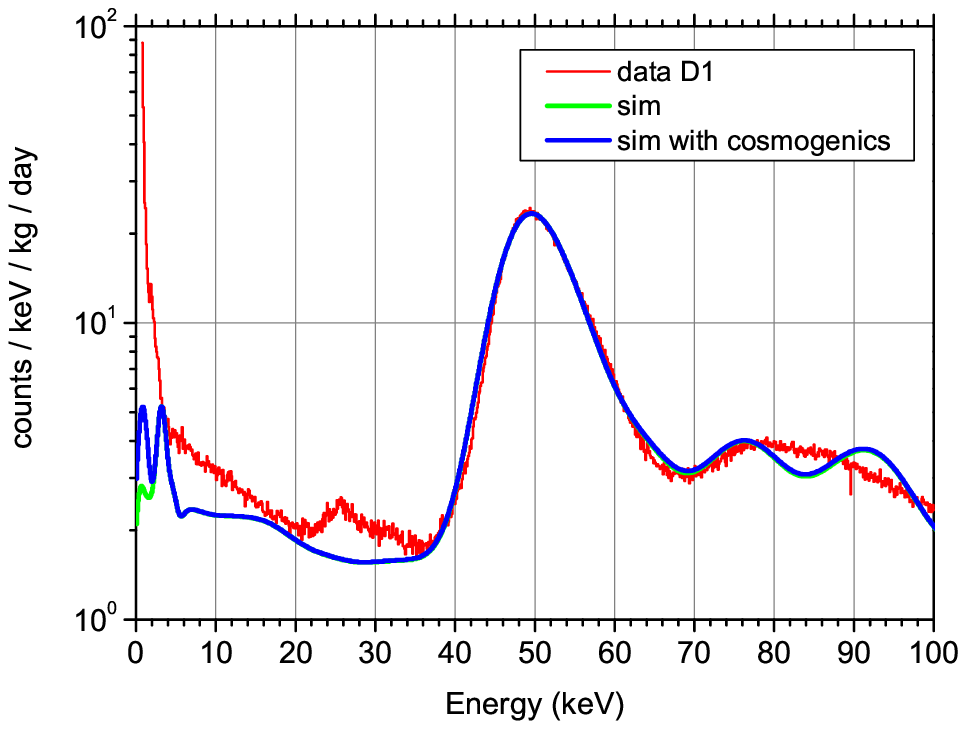}
 \includegraphics[width=0.45\textwidth]{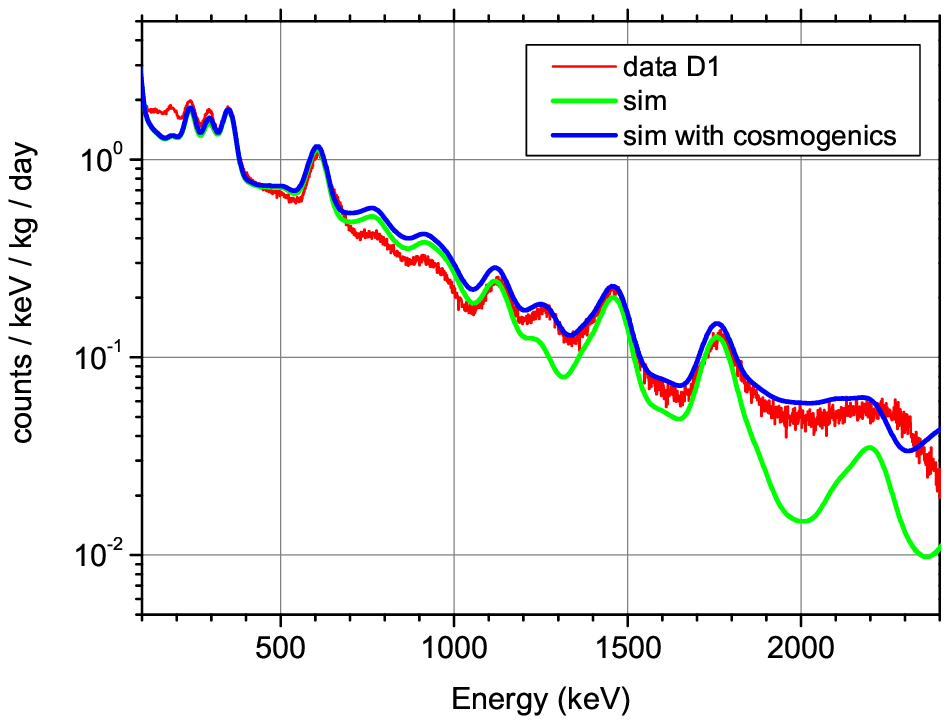}
 \includegraphics[width=0.45\textwidth]{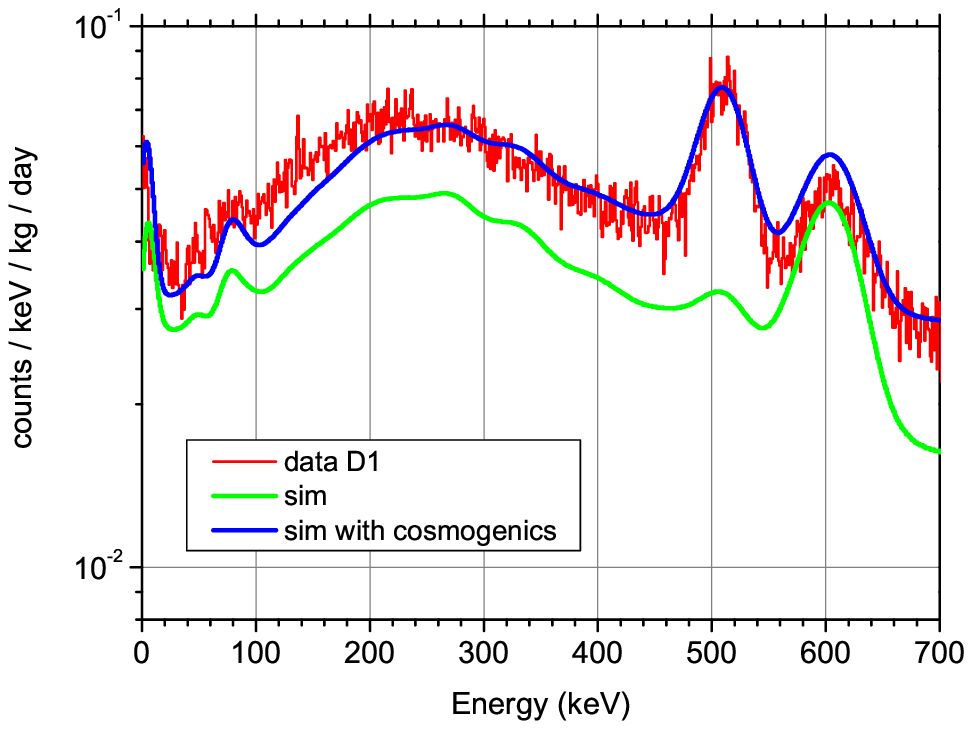}
 \includegraphics[width=0.45\textwidth]{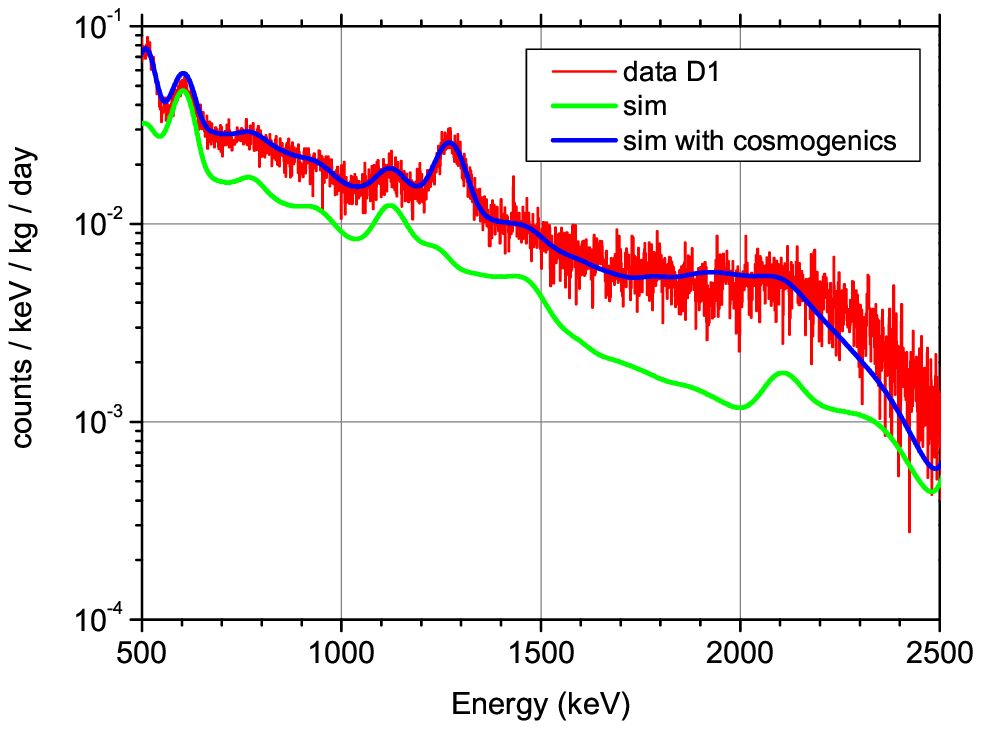}
 \caption{Comparison of the energy spectra summing all the simulated contributions (before and after adding the cosmogenics) with the measured data for ANAIS--25 D1 detector considering anticoincidence (top) and coincidence (bottom) data at low
energy (left) and high energy (right). Spectra for D0 detector are
similar to those of D1.}
  \label{anais25comparison}
\end{figure*}

Figure~\ref{anais25comparison} compares the energy spectrum summing
all the simulated contributions described above with the measured
data for ANAIS--25 detectors, considering anticoincidences or
coincidences between the detectors. The last data taken at the
ANAIS--25 set-up, from June 2014 to March 2015 corresponding to
231.55~d (live time) have been considered here; in these data most
of the cosmogenic isotopes had decayed. A good agreement is obtained
at high energy, but in the very low energy region some relevant
contributions seem to be missing. Since upper limits on radionuclide
activity have been used for several components, the background could
be overestimated in some energy regions. The inclusion of
cosmogenics was essential to reproduce in particular coincidence
data.

The region from about 100 to 200~keV was not well reproduced by
simulations, mainly in the anticoincidence spectrum; the
underestimation is partly solved when adding in the model the
$^{235}$U activity from the PMTs corresponding to the measured
$^{238}$U value and assuming the natural isotopic abundances of
uranium, $\sim$7.2~mBq/PMT. This value is very similar to that
reported at \cite{damapmts}. In addition, the overestimation of the
simulation around 92~keV has been suppressed by reducing the
$^{238}$U upper limit for the copper vessel and quartz windows to
that of $^{226}$Ra\footnote{Upper limits from gamma spectroscopy for
the activity of isotopes at the upper part of the $^{238}$U chain
are typically much larger than those at the lower part starting on
$^{226}$Ra (as it can be seen in table~\ref{extcon}) because of the
very low intensity of the gamma emissions at that chain segment.}.

\begin{figure}
\centering
  \includegraphics[width=0.45\textwidth]{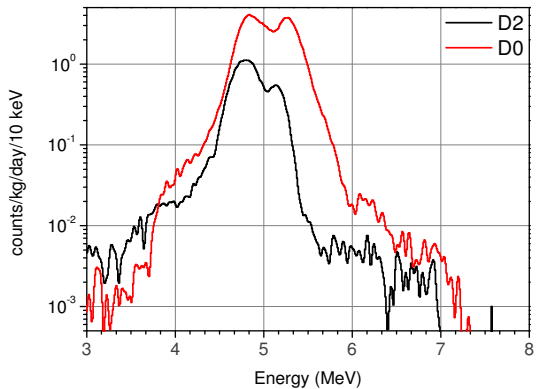} 
 \caption{Alpha region of the measured spectra for D0 (similar to that of D1) and D2 detectors. The peak from $^{210}$Po does not show the expected shape for a bulk contamination.}
  \label{alpharegion}
\end{figure}

\begin{figure}
\centering
  \includegraphics[width=0.45\textwidth]{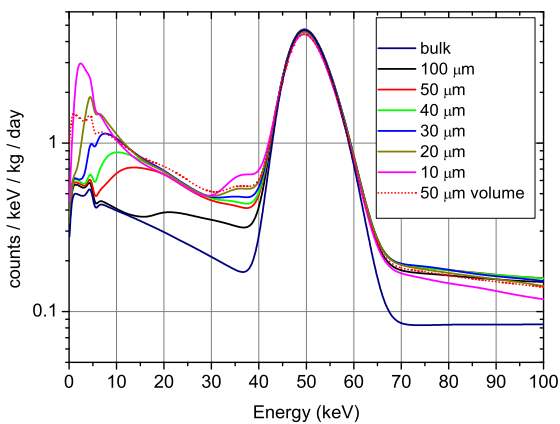} 
 \caption{Comparison of the low energy spectra simulated for $^{210}$Pb emissions from the crystal bulk or from the surface, considering different constant depths all around the crystal; results for uniform emissions from a 50-$\mu$m-thick layer are presented too. All simulations are normalized for an activity of 0.7~mBq/kg.}
  \label{Pb210Surface}
\end{figure}

In an attempt to find out the origin of the missing contributions at
the very low energy region, different hypotheses have been analyzed:
\begin{itemize}
\item It was found that the inclusion in the model of an additional activity of $\sim$0.2~mBq/kg
of $^{3}$H in the NaI crystals significantly improves the agreement
with data at low energy. This value is about twice the upper limit
set for DAMA/LIBRA crystals ($<$0.09~mBq/kg \cite{bernabei2008}),
but lower than the saturation activity which can be deduced from the
production rates at sea level of $^{3}$H in NaI calculated in
\cite{mei} or \cite{mei2}, as described in \cite{anaislrtcosmo}.
\item The unexplained peak around 25~keV could be due to the cosmogenic
production of $^{109}$Cd. This isotope decays by electron capture to
the 88-keV isomeric state of the daughter, having a half-life of
461.9~days, and therefore the peak may correspond to the binding
energy of the K-shell of Ag. The observed peak can be reproduced
using different exposure conditions and production rates of the
order of the estimates made by convoluting production cross-sections
with the cosmic neutron spectrum (see for instance
\cite{dmicethesis}, where a calculated rate of 4.8~kg$^{-1}$d$^{-1}$
is reported). It is worth noting that, being the case, an additional
peak around 3.5~keV (Ag L-shell binding energy) is also expected,
being 5.4 the ratio between K and L-shell EC probabilities
\cite{ddep}.
\item At the alpha region of the energy spectra measured for the detectors, the prominent peak
due to the $^{210}$Po emission (at the decay sequence of $^{210}$Pb)
does not show the pure structure expected from a crystal bulk
$^{210}$Pb contamination (see figure~\ref{alpharegion})
\cite{Clarathesis}. Therefore, the possibility of a surface
deposition was carefully analyzed. Figure~\ref{Pb210Surface}
compares the low energy spectra simulated assuming $^{210}$Pb in
bulk or in surface; since the surface contamination profile is
unknown, different constant depths all around the crystal have been
considered\footnote{Simulations with uniform emissions from the
whole volume of a surface layer of the crystal have been also made;
as an example, results for $^{210}$Pb contamination distributed in a
50-$\mu$m-thick surface layer are shown in figure~\ref{Pb210Surface}
too.}. The alpha emission is fully absorbed and only a small
continuum appears at the left side of the peak for depths at or
below 30~$\mu$m in the simulation\footnote{The range of a 5-MeV
alpha particle in NaI is 29 $\mu$m, following NIST data
\cite{nist}.}. The need to reproduce the low energy region of the
measured spectrum dominated by $^{210}$Pb emissions excludes surface
contaminations at a very small depth. The best option to reproduce
the whole range of data was found when considering half of the
$^{210}$Pb content in bulk and the other half on surface from a
constant depth of 100~$\mu$m; the bulk and surface proportion was
fixed following the observed almost symmetric double structure of
the $^{210}$Po peak for D0 and D1 detectors. This result has to be
taken very cautiously, energy conversion into visible signal is
assumed to be constant throughout the crystal in our simulation, but
there should be a difference between energy depositions for alpha
particles in two regions, as we are indirectly assuming when
interpreting the double structure in the alpha peak as due to
$^{210}$Po. These two regions could be surface and bulk. However,
simple tests, including a similar reduced energy conversion for
beta/gamma energy depositions having the same distribution than
alpha contamination can be discarded, because it affects so much at
the low energy events that it should have been clearly observed in
the data. Nevertheless, we cannot discard some spatial dependence of
the energy conversion that could affect differently alpha particles
than beta/gammas and then, it could affect too the energy
depositions at the lowest energies (below 10~keV and not in the
50~keV energy scale).
\end{itemize}

\begin{figure*}
\centering
  \includegraphics[width=0.45\textwidth]{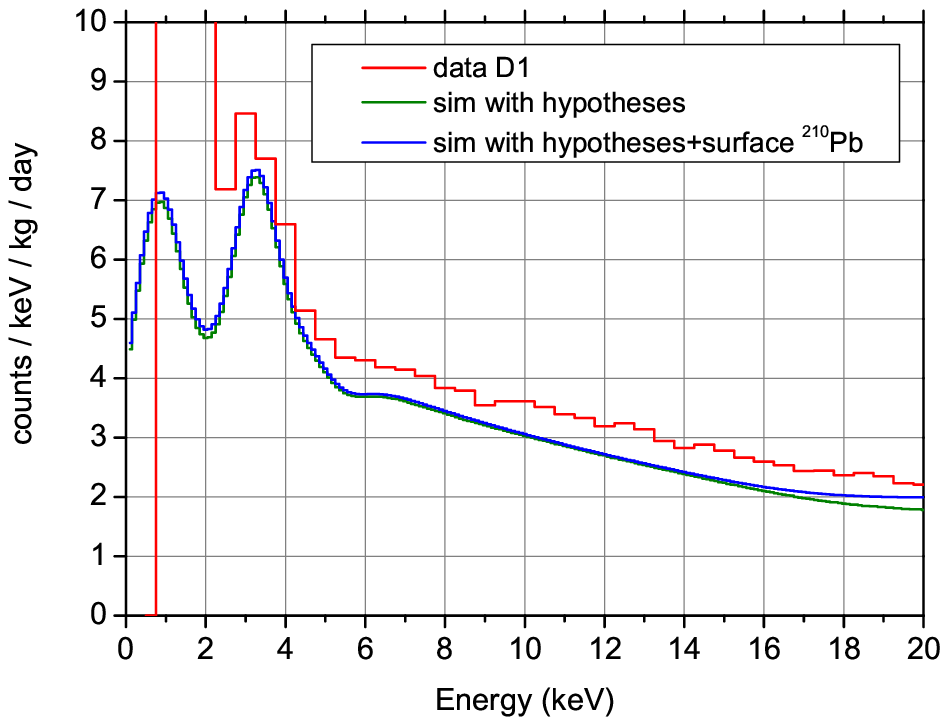}
  \includegraphics[width=0.45\textwidth]{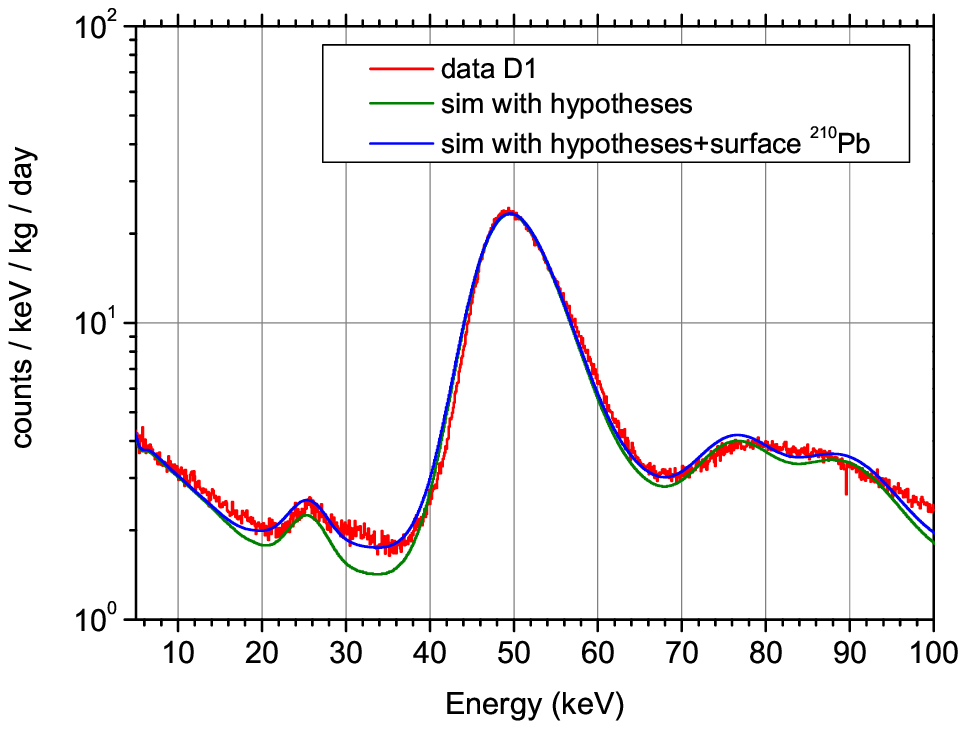}
  \includegraphics[width=0.45\textwidth]{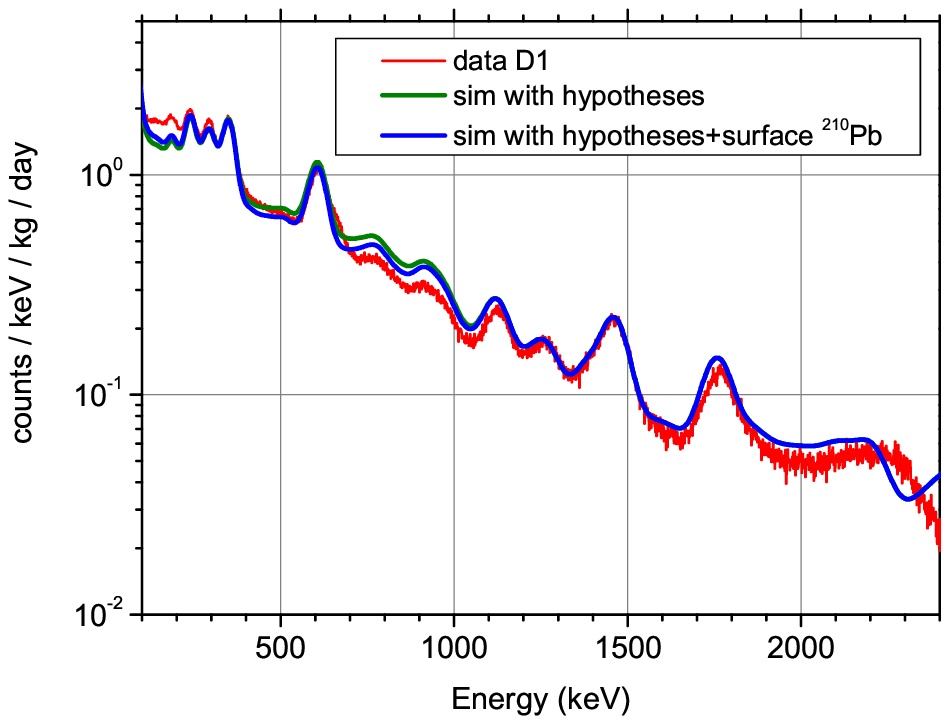}
 \caption{Effect of the consideration of some background hypotheses in the
spectra of ANAIS--25 detectors at all energy ranges for
anticoincidence data (see text). The inclusion of some reduced
$^{238}$U upper limits, $^{235}$U at PMTs and $^{3}$H and $^{109}$Cd
at crystals has been considered (green line); the additional
assumption of half of the $^{210}$Pb emission from a depth of
100~$\mu$m on the crystal surface is separately shown (blue line).
The considered hypotheses allow to significantly improve the overall
agreement with measured data.}
  \label{anais25hypothesis}
\end{figure*}

Figure~\ref{anais25hypothesis} compares again the measured ANAIS--25
spectra with the simulated ones including all these hypotheses, for
anticoincidence data. The model still gives an overestimation from
0.6 to 1~MeV, since upper limits have been considered for several
contaminations, and an underestimation from 100 to 200~keV. The
inclusion of both $^{3}$H and $^{109}$Cd contributions significantly
improves the agreement in the lowest energy region. The additional
inclusion of $^{210}$Pb partly on surface helps to reproduce the
region from 30 to 40~keV.

\begin{figure*}
\centering
 \includegraphics[width=0.45\textwidth]{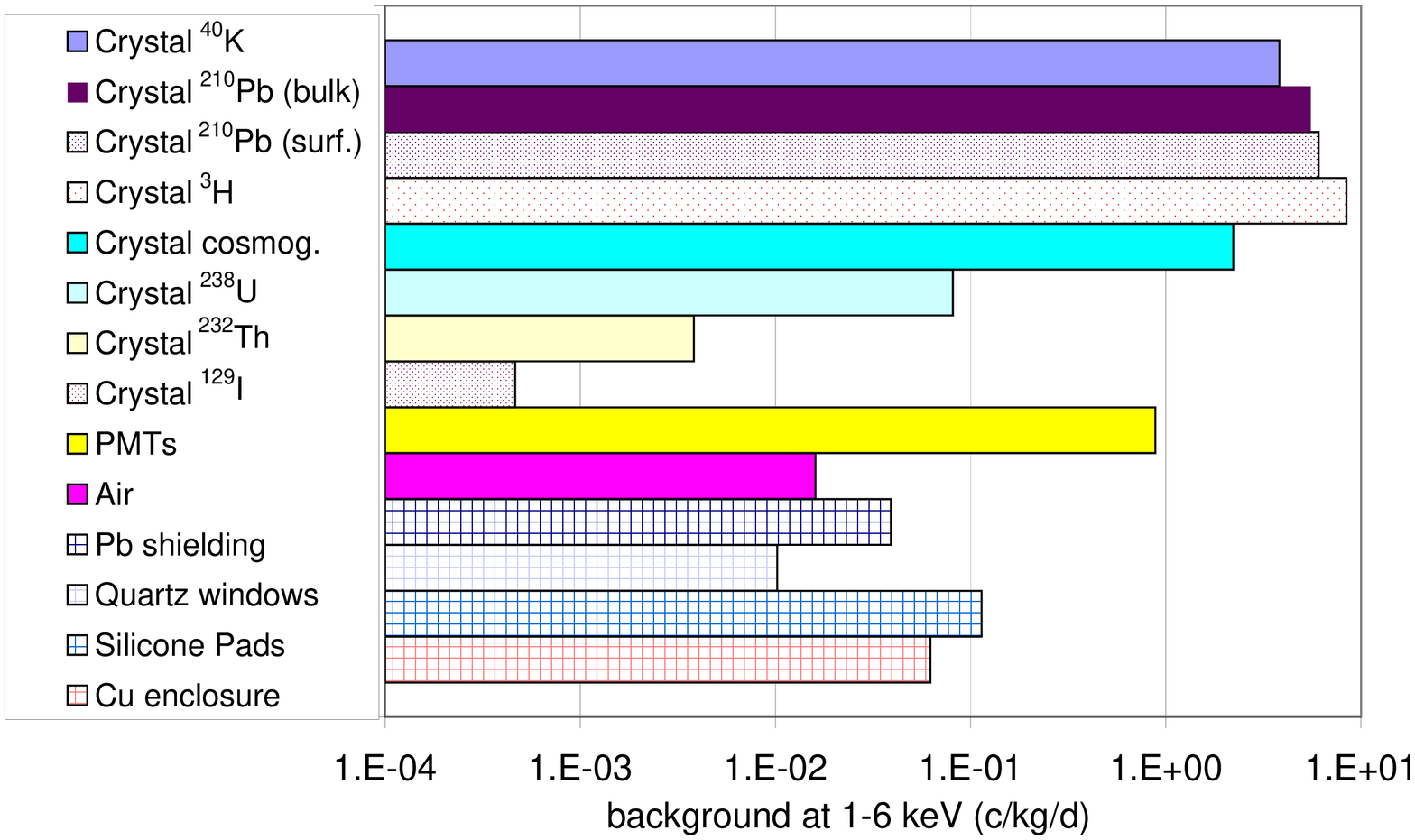}
  \includegraphics[width=0.45\textwidth]{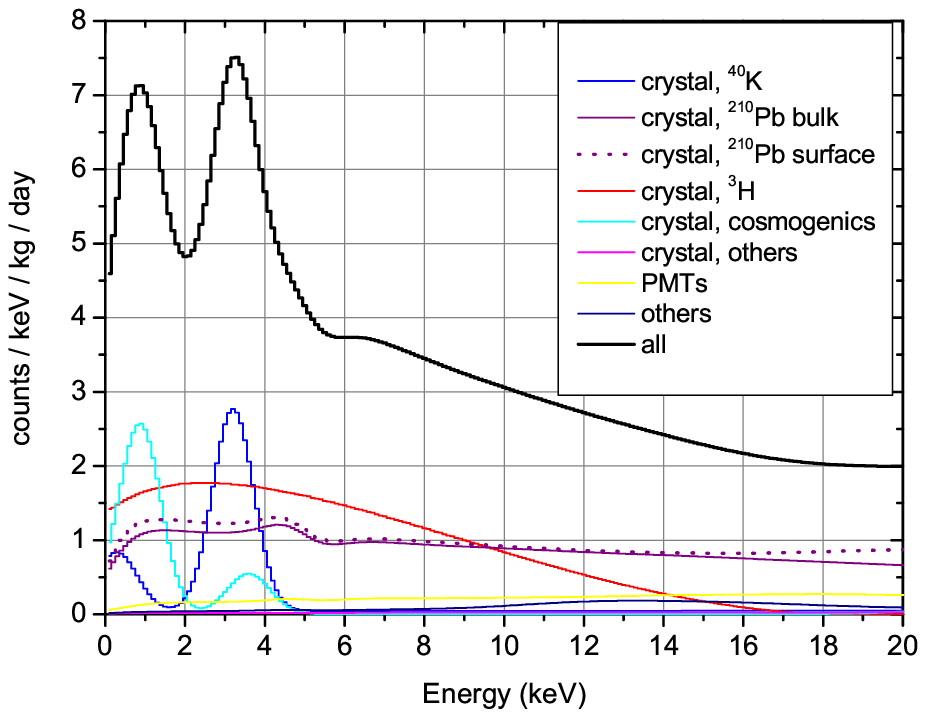}
 \caption{ANAIS--25 background model: expected rates from different background sources in the region of 1-6 keVee (left) and the corresponding spectra at the very low energy
region (right). In the left plot, some contributions have been
estimated from a directly quantified activity (filled bars) but
others from upper limits (plaid bars) or hypothesized activities
(dotted bars).}
  \label{anais25model}
\end{figure*}

Figure~\ref{anais25model}, left summarizes the different
contributions from the explained background model of ANAIS--25
detectors, for anticoincidence data, to the rate in the region from
1 to 6~keVee. The energy spectra expected from different background
sources in the very low energy region for anticoincidence data are
plotted in figure~\ref{anais25model}, right, together with the sum
of all contributions. In the RoI $^{210}$Pb and $^{3}$H continua and
$^{40}$K, $^{22}$Na and $^{109}$Cd peaks are the most significant
contributions.

\subsection{D2 module at ANAIS--37}

\begin{figure*}
\centering
 \includegraphics[width=0.45\textwidth]{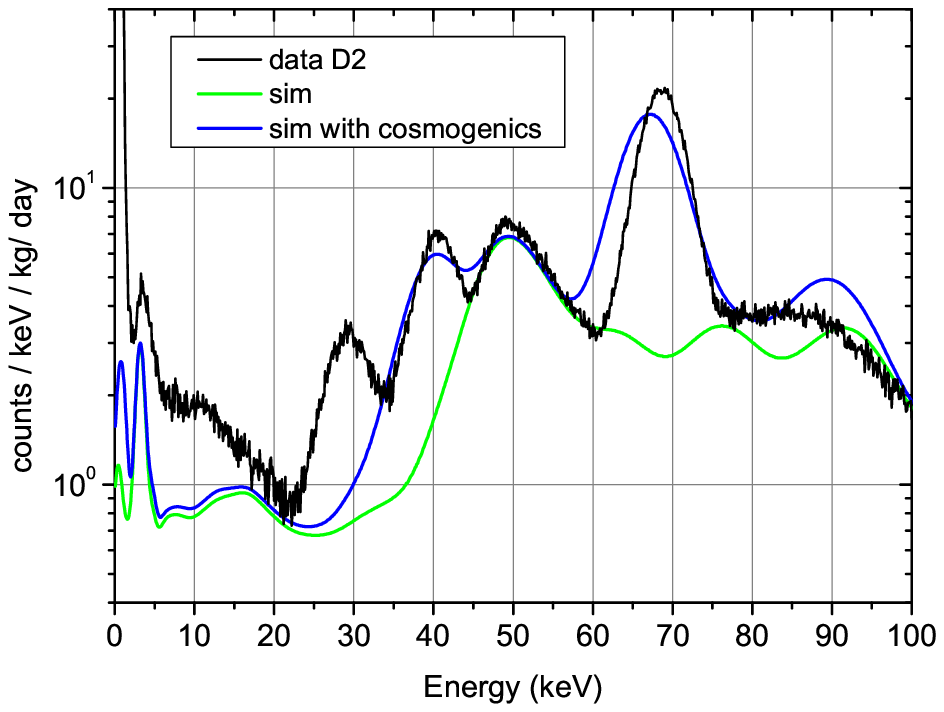}
 \includegraphics[width=0.45\textwidth]{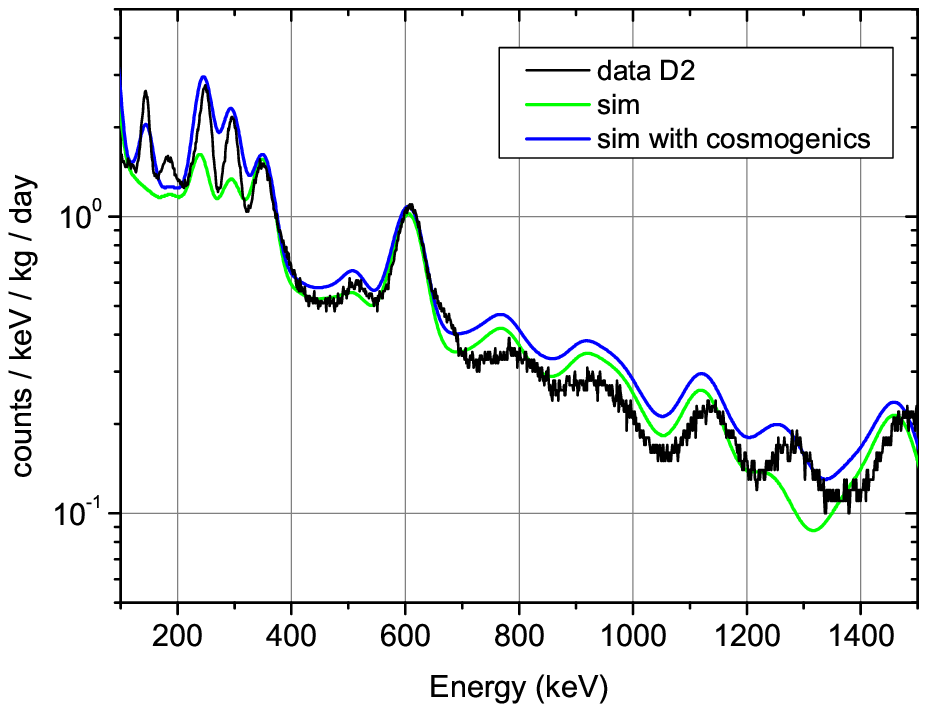}
 \includegraphics[width=0.45\textwidth]{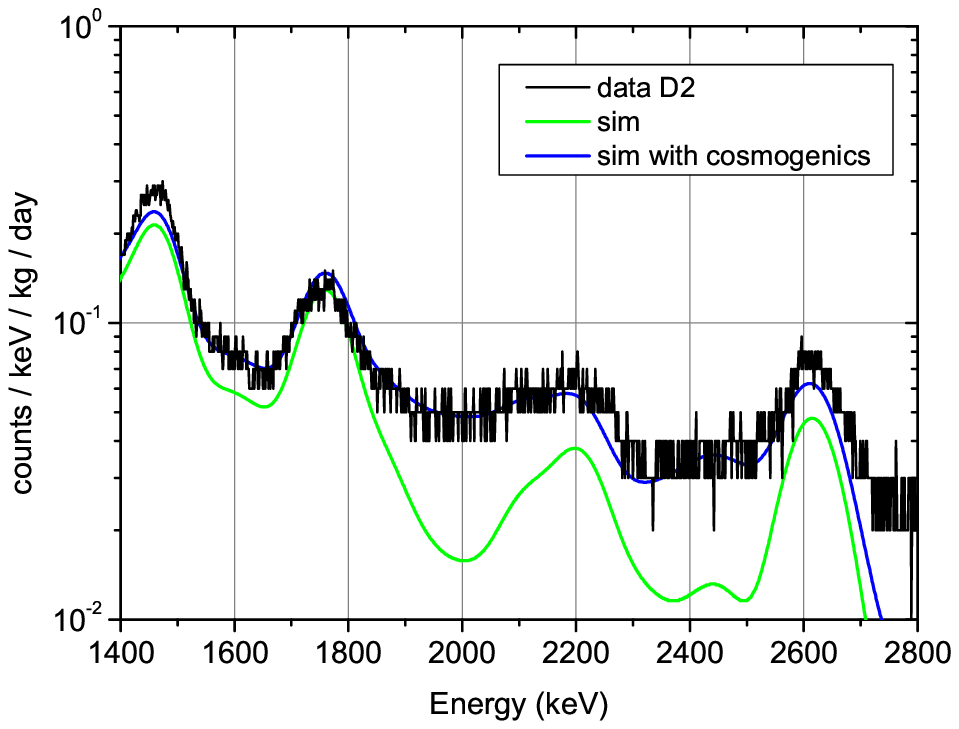}
 \caption{Comparison of the energy spectrum measured for D2 detector at the ANAIS--37 set-up with the corresponding simulation summing all contributions (before and after adding the cosmogenics) at
 low, medium and high energy regions. Anticoincidence data are shown in the low energy region.}
  \label{D2comparison}
\end{figure*}

Figure \ref{D2comparison} compares the energy spectrum measured for
D2 detector at the ANAIS--37 set-up with the corresponding
simulation. The data taken with this set-up from May to September
2015 for 89.5~d have been considered here; in these data the
cosmogenic activation has significantly, but not completely,
decayed. For the simulation, the input activities given in tables
\ref{extcon} and \ref{intcon} for D2 have been assumed. The initial
activity of I and Te products cosmogenically induced in the NaI
crystal has been taken the same determined for D0 and D1 detectors,
although previous exposure history of the detectors is not the same
and some differences could be expected in other longer half-life
isotopes, for I and Te products saturation should have been reached
nonetheless. For $^{22}$Na, the initial activity quantified
specifically for D2 detector (see section \ref{sources}) has been
considered. The overall agreement between data and simulation in
figure~\ref{D2comparison} is quite satisfactory except for some
energy regions. As for ANAIS--25 detectors, $^{210}$Pb emissions at
low energy are well reproduced assuming the activity deduced from
the alpha rate. At medium energies around 1~MeV, the simulation is
clearly overestimated; this could be due again to the fact that
upper limits on radionuclide activity have been used for several
components.

The observed peak around 185~keV in D2 data and not reproduced by
simulation (as well as the underestimation of that at $\sim$145~keV)
could be justified by the content of $^{235}$U from the PMTs, also
considered for D0 and D1 detectors (see section~\ref{secanais25}).
Again as for ANAIS--25, the overestimation of the simulation around
92~keV can be partially suppressed by reducing the $^{238}$U upper
limit for the copper vessel and quartz windows to that of
$^{226}$Ra. Some plausible hypotheses have been analyzed to find an
explanation to the relevant discrepancies in the low energy region:
\begin{itemize}
\item The prominent line registered at $\sim$28~keV is not explained
by the cosmogenic isotopes identified in ANAIS--25 data. $^{113}$Sn,
having a half-life of 115.1~days and decaying by electron capture
mainly to a 391.7~keV isomeric state of the daughter, could justify
this spectral feature as binding energy of In K-shell and activation
of the isotope is possible. As for $^{109}$Cd, the observed peak can
be reproduced using different exposure conditions and production
rates of the order of the estimates made by convoluting production
cross-sections with the cosmic neutron spectrum (at
\cite{dmicethesis}, a calculated rate of 9~kg$^{-1}$d$^{-1}$ is
reported and a measured value of 16~kg$^{-1}$d$^{-1}$ presented). A
peak around 4~keV is expected from $^{113}$Sn as the binding energy
of the L-shell of In (the ratio between the probabilities of
electron capture for K and L shells is 7.4 for decay to the isomeric
state \cite{ddep}), but since the half-life of $^{113}$Sn is not too
large, it should not be a problem in the long term. $^{113}$Sn
induced in D0/D1 detectors should have decayed in the data analyzed
here. No hint of the presence of $^{109}$Cd, as seen in ANAIS--25
detectors, can be observed for the moment in D2 data; as for
$^{22}$Na, the initial activity of this isotope, having a longer
half-life than $^{113}$Sn, could be lower than in D0 and D1 due to
the different time spent at Colorado.
\item The spectrum shape and rate observed below 20~keV in D2 data
are not completely reproduced by simulation. Several possibilities
have been explored:
\begin{itemize}
\item Considering the amount of $^{3}$H deduced for ANAIS--25 detector,
the spectral shape of its beta emission does not fully explain the
observed background below 20 keV; therefore, its content has been
fixed just to the upper limit set by DAMA/ LIBRA (0.09~mBq/kg
\cite{bernabei2008}). Due to the shorter exposure to cosmic rays in
Colorado of D2 in comparison to D0 and D1, a lower $^{3}$H activity
is expected.
\item The possibility of $^{210}$Pb emissions on the crystal surface
instead of in bulk has been deeply studied too. As shown in
figure~\ref{alpharegion}, the double structure of the $^{210}$Po
peak observed at the alpha region of the energy spectra of D2
detector is more asymmetric than that measured for D0/D1; according
to this peak structure, a fourth of the $^{210}$Pb contamination has
been considered in the crystal bulk and the rest on surface. For D2
detector, and considering all the possibilities analyzed (see
figure~\ref{Pb210Surface} and section~\ref{secanais25} for more
details), the best option to reproduce the low energy region of the
spectra dominated by $^{210}$Pb emissions was a surface
contamination from a constant depth of 30~$\mu$m.
\end{itemize}
\end{itemize}

\begin{figure*}
\centering
 \includegraphics[width=0.44\textwidth]{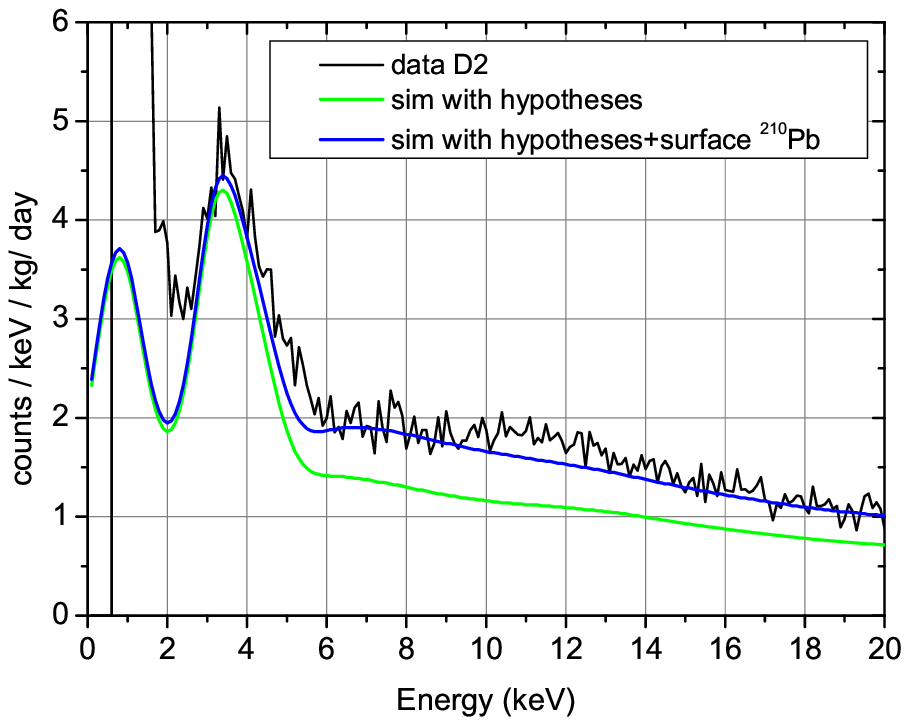}
 \includegraphics[width=0.45\textwidth]{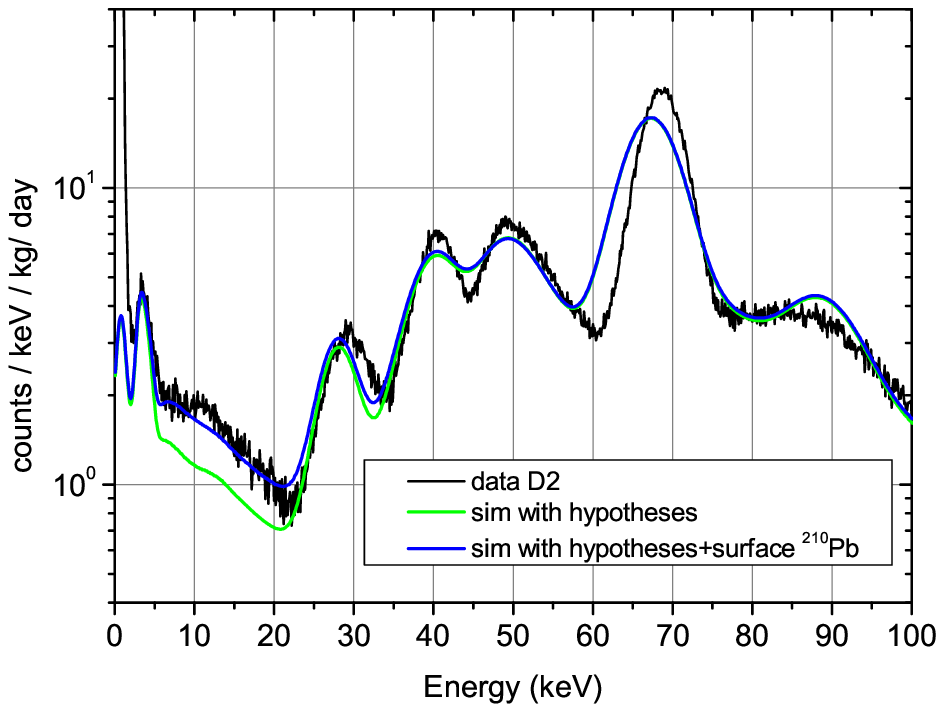}
 \includegraphics[width=0.45\textwidth]{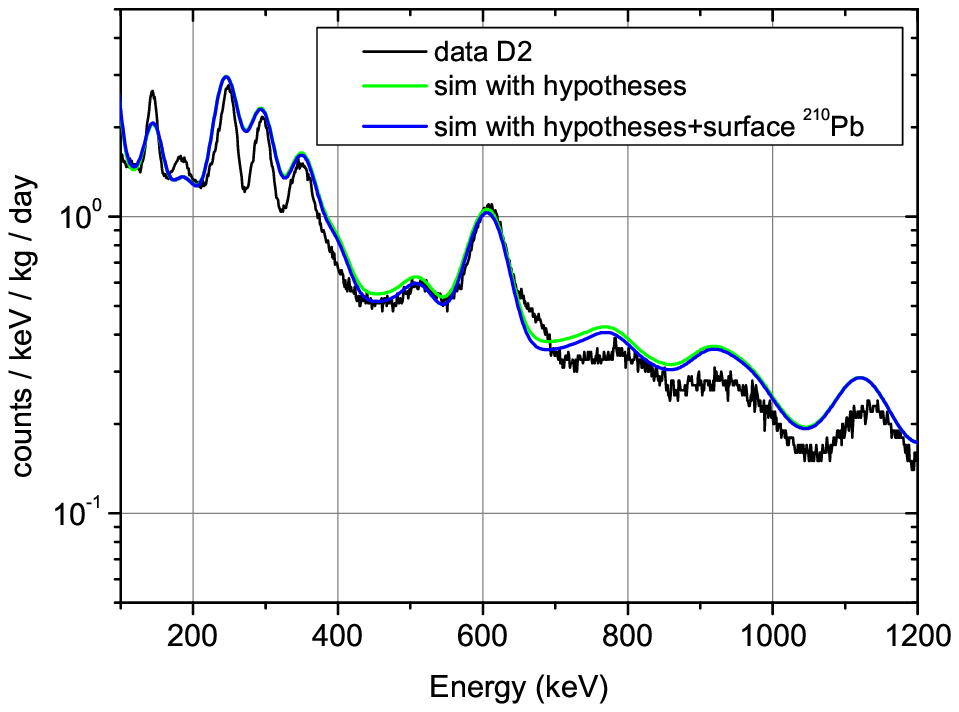}
 \caption{Effect of the consideration of the plausible analyzed background hypotheses (see text) in the spectra of D2 detector at different energy ranges (at the highest energies the hypotheses have no effect).
 The inclusion of some reduced $^{238}$U upper limits, $^{235}$U at
PMTs and $^{3}$H and  $^{113}$Sn at crystal has been considered
(green line); the additional assumption of part of the $^{210}$Pb
emission from a depth of 30~$\mu$m on the crystal surface is
separately shown (blue line). The considered hypotheses are
essential to improve the overall agreement with measured data.}
  \label{D2hypothesis}
\end{figure*}

Figure \ref{D2hypothesis} compares the measured D2 spectra with the
simulated ones taking into consideration all the described
hypotheses, which significantly improve the agreement. It is
remarkable that the inclusion of $^{3}$H in the model is necessary,
since only surface and bulk $^{210}$Pb emissions cannot reproduce
the registered spectrum. The higher continuum level below 40 keV
produced by the $^{210}$Pb surface emission in comparison with the
bulk contamination helps to make the model be closer to the measured
data.

\begin{figure*}
\centering
 \includegraphics[width=0.45\textwidth]{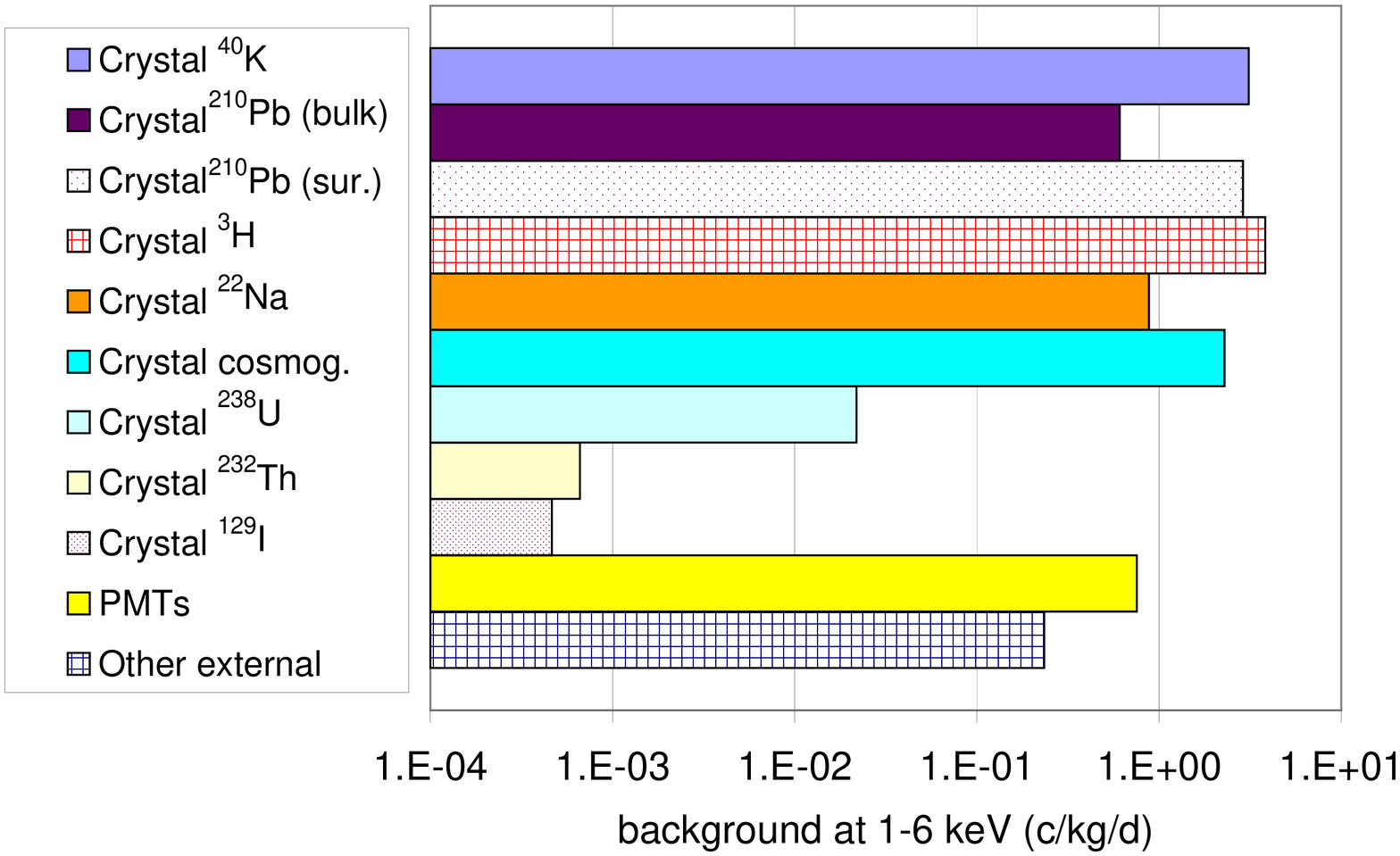}
  \includegraphics[width=0.45\textwidth]{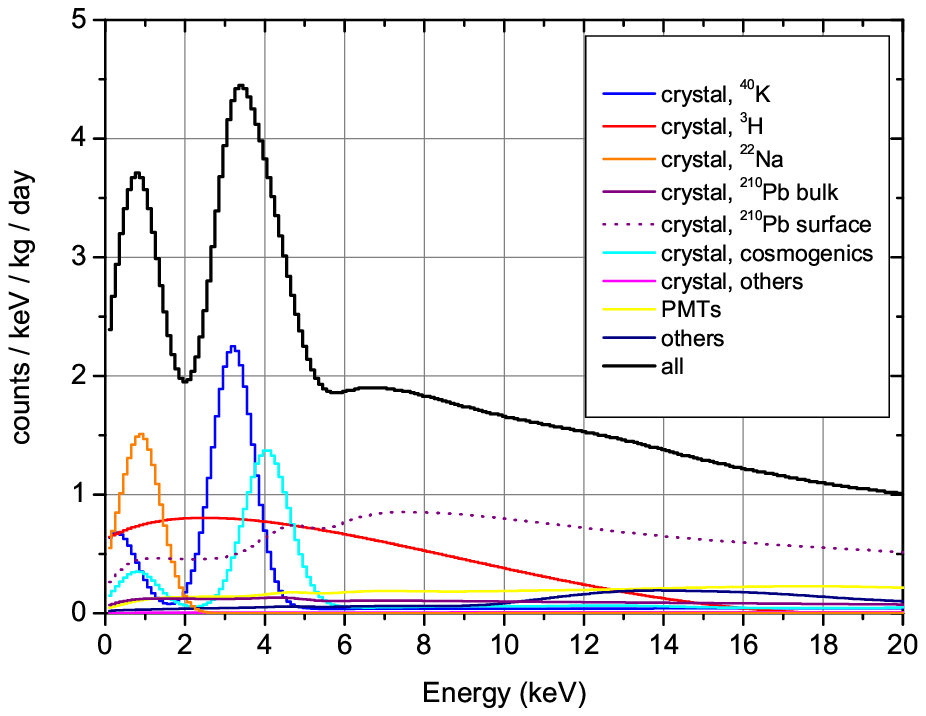}
 \caption{Background model for D2 at the ANAIS--37 set-up: expected rates from different background sources in the region of 1-6~keVee (left) and the corresponding spectra at the very low energy
region (right). In the left plot, some contributions have been
estimated from a directly quantified activity (filled bars) but
others from upper limits (plaid bars) or hypothesized activities
(dotted bars). Contribution from cosmogenic isotopes has been
evaluated for the period corresponding to the real data taking.}
  \label{D2model}
\end{figure*}

Figure \ref{D2model}, left summarizes the different contributions
from the proposed background model of D2 at the ANAIS--37 set-up,
for anticoincidence data, to the rate in the region from 1 to 6
keVee. The energy spectra expected from different background sources
in the very low energy region for anticoincidence data are plotted
in figure~\ref{D2model}, right, together with the sum of all
contributions. In this RoI it has been verified that $^{210}$Pb
contribution has been significantly reduced in comparison to D0 and
D1; but it is worth pointing out that emissions from surface make a
larger contribution than from bulk. Peaks from $^{113}$Sn and, to a
lesser extent, $^{22}$Na will be decreasing in the next future.

As conclusion, and according to our background model, surface
contamination in $^{210}$Pb is needed to reproduce low energy
spectra of D0/D1 and D2 modules, even though in a different amount
and in different depth profile. Although some important assumptions
are required to estimate the contribution of surface contaminants to
the energy spectra, and then, the contaminations depths for
$^{210}$Pb derived from our analysis should not be firmly taken as
stated, we can conclude that a fraction of the total contamination
in $^{210}$Pb could be due to the treatment of the surfaces while
the building of D0/D1 and D2 modules. In \cite{NaIsurface,cuoreRn},
diffusion lengths for radon-induced surface contamination which
could happen due to exposure to air during assembly phase and
storage of detectors are estimated below 1~$\mu$m, depending on
radon concentration, exposure time and features of material surface;
the required depths deduced for ANAIS crystals cannot then be
explained by diffusion from radon surface deposition, but they would
point to other mechanisms allowing the contamination from radon (or
even directly $^{210}$Pb) at the production phase of the crystals or
at the treatment of surfaces. This issue is being further
investigated in collaboration with AS company.

\section{Background projections} \label{projections}

\begin{figure*}
\centering
 \includegraphics[height=0.27\textheight]{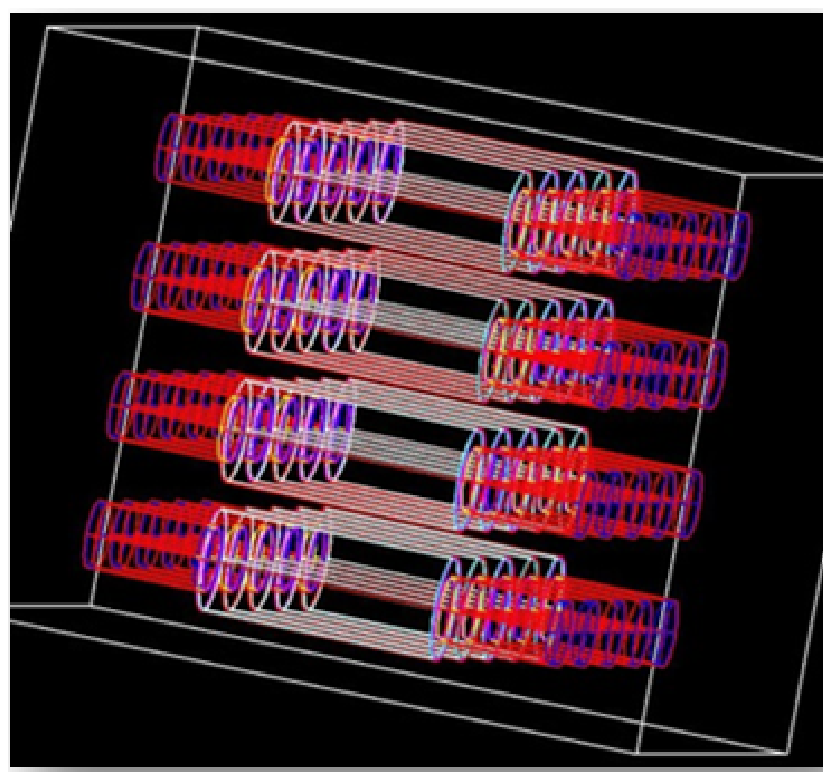} 
 \includegraphics[height=0.27\textheight]{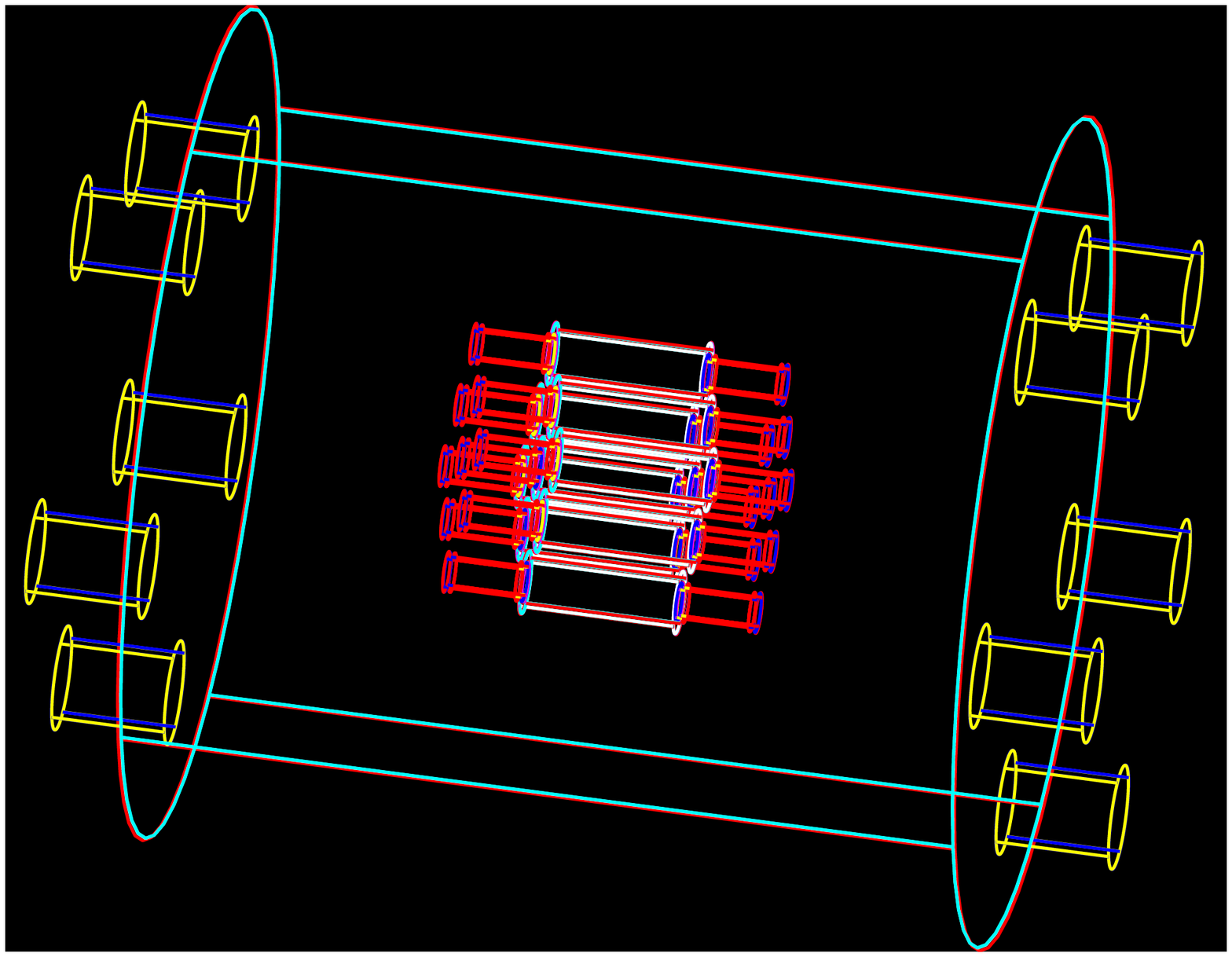} 
 \caption{Geometry of some possible full set-ups for ANAIS, as implemented in Geant 4 simulations: with 4$\times$5 NaI(Tl) modules (left) and with 3$\times$3 NaI(Tl) detectors inside a liquid scintillator veto (right).}
  \label{geometrys2}
\end{figure*}

ANAIS in collaboration with AS is strongly pursuing an improvement
of the radiopurity of the subsequent 12.5~kg modules, trying to
focus in further reduction of the powder $^{40}$K content, in
reducing $^{22}$Na activation by using convenient shieldings while
storing at surface, and avoiding $^{210}$Pb contamination.
In parallel, background rejection power of coincidences has been
analyzed in different experimental scenarios for ANAIS: present
configuration with 3$\times$3 modules (112.5~kg) and also an
enlarged experiment using 4$\times$5 modules (250~kg). An additional
Liquid Scintillator Veto (LSV) surrounding the NaI(Tl) detectors has
been also considered (see figure~\ref{geometrys2}).

Shape and size of ANAIS--37 modules have been assumed in all this
analysis. For the LSV system, Linear AlkylBenzene (LAB) has been
considered as scintillator medium inside a cylindrical stainless
steel vessel read by five 8'' Hamamatsu R5912 PMTs at each side; a
scintillator thickness of 60~cm at top, bottom and side of the
cylindrical vessel has been assumed (corresponding to a total mass
of 3.8~t). For all the full ANAIS set-ups simulated, measured
primordial activities from PMTs and crystals (assuming 0.7~mBq/kg of
bulk $^{210}$Pb and 1.25~mBq/kg of $^{40}$K) have been included. The
cosmogenic initial activity of $^{22}$Na and possible tritium
contribution have been taken into account too, at the levels deduced
for D2. Concerning the activity of the LSV system itself, values
from the literature have been considered: from NEXT for stainless
steel \cite{next}, SNOlab for PMTs \cite{snolab} and JUNO for LAB
scintillator \cite{juno}. According to the obtained results for the
assumed geometry and activities, the contribution to the background
levels of the LSV components is negligible (two orders of magnitude
lower) in comparison to that of the NaI(Tl) crystals; for this
reason, the 3$\times$3 modules set-up without a LSV system has not
been independently simulated.

\begin{figure}
\centering
 \includegraphics[width=0.45\textwidth]{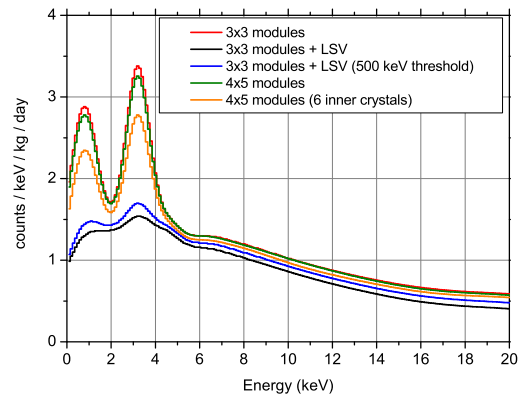} 
 \caption{Energy spectra at the very low energy region expected
from all the considered background sources in different set-ups and
anticoincidence conditions (see text). The presently achieved
radiopurity of D2 module has been assumed for all the modules.}
  \label{veteff}
\end{figure}

\begin{figure}
\centering
 \includegraphics[width=0.45\textwidth]{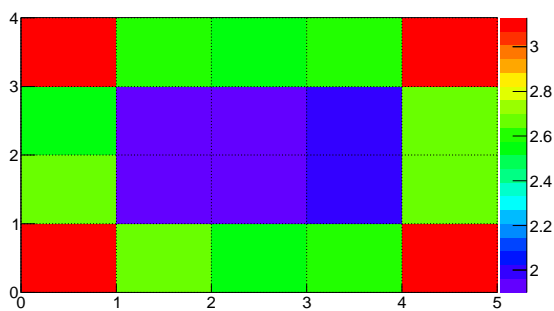} 
 \caption{Distribution of the background level (expressed in c/kg/d) registered in the 1-6~keVee region in anticoincidence at each
 crystal over a matrix of 4 (vertical) by 5 (horizontal) detectors, for $^{40}$K emissions from the NaI(Tl) crystals. Similar plots are obtained for $^{22}$Na also from crystals and for PMTs emissions; they indicate that the analysis of only the six inner modules could
be advantageous.}
  \label{distributionK40}
\end{figure}

Figure~\ref{veteff} compares the energy spectra at the very low
energy region expected from all the considered background sources in
different situations, to illustrate the effect of anticoincidence
between crystals and also with the LSV. Spectra have been evaluated
for the 3$\times$3 detectors set-up considering anticoincidence only
between crystals, also with an ideal LSV and with a LSV having a
500~keV energy threshold. Spectra obtained in anticoincidence at the
4$\times$5 modules set-up are also presented; in this configuration
it has been observed that considering only the six inner crystals,
and not all of the twenty detectors, produces a lower background
level at the RoI (see figure~\ref{distributionK40}) and results in
these two conditions are shown. As expected, the reduction by
anticoincidence is more modest in a 3$\times$3 matrix than in a
4$\times$5 configuration. But the implementation of a LSV system at
the 3$\times$3 matrix set-up is more effective that the increase of
NaI modules, for fixed radiopurity levels. Concerning the
contribution at the RoI from $^{210}$Pb in crystals, it is worth
noting that it could be higher in case of a surface contamination
(as shown in figure~\ref{Pb210Surface}); for instance, if all the
$^{210}$Pb was on surface at a constant depth of 30~$\mu$m, the
increase in the rate in the 1-6~keVee region in comparison to a pure
bulk contamination would be of 1.6.

Table~\ref{redfac} summarizes the reduction factors (R.F.) obtained
in the considered situations at the 1-6~keVee energy region, for all
the contributions altogether and separately for some background
sources, those for which high reduction is expected: $^{40}$K and
$^{22}$Na at crystals and PMT contaminations. The R.F. is computed
from the ratio between the rates in the 1-6~keVee after coincidence
events rejection and before any rejection and therefore corresponds
to the percentage of remaining background. The veto by the LSV is
very effective for the $^{40}$K and $^{22}$Na peaks, but since it is
useless for the other relevant background sources at the RoI
($^{3}$H and $^{210}$Pb) the estimated overall R.F. is in the end of
55.6\% for the ideal LSV and 59.1\% assuming a 500~keV threshold
(see table~\ref{redfac}). For the assumed background model, about
70\% of the background obtained for the 3$\times$3 crystals
configuration in anticoincidence remains after the vetoing effect of
the LSV.

\begin{table*}
\caption{Reduction factor (R.F., defined as \% of background
remaining after anticoincidence in the 1-6~keVee region) computed
for several set-ups and applying anticoincidence rejection only
between crystals or also with the LSV (see text). It has been
evaluated separately for some background sources and for the overall
background in different experimental scenarios.} \label{redfac}
\begin{center}
\begin{tabular}{lcccc}
\hline\noalign{\smallskip} Set-up &  R.F.(\%) & R.F.(\%)
& R.F.(\%) & R.F.(\%) \\
& $^{40}$K from crystals & $^{22}$Na from crystals & PMTs & Total \\
\noalign{\smallskip}\hline\noalign{\smallskip}

3$\times$3 modules &  69.0  &  62.4 & 62.3 & 83.7 \\

3$\times$3 modules + LSV &   11.9 & 1.2 & 7.3 &  55.6 \\

3$\times$3 modules + LSV (500 keV threshold) &  15.5 & 5.7 & 29.3 & 59.1 \\

4$\times$5 modules  & 62.7 & 55.1 & 45.9 & 77.9 \\

4$\times$5 modules (6 inner crystals)  &   48.6 & 39.8 & 32.5  & 70.3\\
\noalign{\smallskip}\hline
\end{tabular}
\end{center}
\end{table*}


\section{Conclusions} \label{conclusions}

The background models constructed for D0 and D1 detectors at the
ANAIS--25 set-up and for D2 module at ANAIS--37 provide a good
description of measured data at all energy ranges and at different
analysis conditions (coincidence/anticoincidence). The measured
activity in external components and in crystal, including cosmogenic
products and quantified combining different analysis techniques,
roughly explains the observed background. But the inclusion of some
additional hypotheses, like the presence of cosmogenic isotopes
which cannot be directly quantified or partial crystal surface
$^{210}$Pb contamination, significantly improves the agreement
between model and real data.

Reduction factors for the background at the very low energy region
thanks to the rejection of coincident events have been computed for
the full ANAIS set-up and for several hypothetical scenarios, like a
matrix of NaI(Tl) crystals corresponding to a total mass of 250~kg,
or the use of a liquid scintillator veto in the 3$\times$3 modules
configuration.

The measured background in D2 is already about 2~counts/(keV~kg~d)
at $\sim$6~keVee but reduction is still possible thanks to the
increase of the background rejection power in a detector matrix
set-up and an improved control of radiopurity. Thanks to the already
achieved reduction in D2 detector of $^{210}$Pb activity, its
contribution in the RoI is at some tenths of count/(keV~kg~d). That
of PMTs and other external components is below
0.2~counts/(keV~kg~d). $^{40}$K and $^{22}$Na peaks can be
significantly reduced by anticoincidence operation of the full
experiment. With the radiopurity levels achieved in D2 detector, a
background rate below 2~counts/(keV~kg~d) above 4~keVee is expected
for a 3$\times$3 matrix of crystals amounting to 112.5~kg of mass;
with these activity levels of D2, the use of a LSV is clearly a
better alternative than the increase of the total mass up to 250~kg,
and the background rate is reduced below 1.5~counts/(keV~kg~d) at
the whole RoI. In this situation, the fully-absorbed emission from
$^{3}$H could become the main contribution in the region of
interest; therefore, a shielding against cosmogenic activation has
been procured for the production of the new ANAIS NaI(Tl) crystals.
Additionally, a further reduction of $^{210}$Pb and $^{40}$K in
Alpha Spectra detectors of WIMPScint-III grade can be foreseen
thanks to improved purification and surface machining protocols.

\begin{acknowledgements}
This work has been supported by the Spanish Ministerio de
Econom\'{i}a y Competitividad and the European Regional Development
Fund (MINECO-FEDER) (FPA2011-23749 and FPA2014- 55986-P), the
Consolider-Ingenio 2010 Programme under grants MULTIDARK CSD2009-
00064 and CPAN CSD2007-00042, and the Gobierno de Arag\'{o}n and the
European Social Fund (Group in Nuclear and Astroparticle Physics).
P. Villar is supported by the MINECO Subprograma de Formaci\'{o}n de
Personal Investigador. We also acknowledge LSC and GIFNA staff for
their support.
\end{acknowledgements}



\end{document}